\pdfoutput=1

\documentclass{aa}  
\usepackage{graphicx}
\usepackage{natbib}
\usepackage[figuresright]{rotating}
\usepackage{amsmath}
\usepackage{amssymb}
\usepackage{bbm}
\bibpunct{(}{)}{;}{a}{}{,}
\usepackage{txfonts}
\begin{document}

\title{Comparative clustering analysis of variable stars in the 
Hipparcos,
OGLE Large Magellanic Cloud and CoRoT\thanks{The CoRoT
space mission was developed and is operated by the French space agency CNES,
with participation of ESA's RSSD and Science Programmes, Austria, Belgium,
Brazil, Germany, and Spain.} exoplanet databases}

\author{L. M. Sarro\inst{1,2} \and J. Debosscher\inst{3} \and
          C. Aerts\inst{3,4} \and M. L\'opez\inst{2} }
          
\institute{Dpt.\ de Inteligencia Artificial , UNED, Juan del Rosal, 16, 28040
     Madrid, Spain 
\and Spanish VO - LAEX-CAB (INTA-CSIC), Postal
     address.- LAEFF, European Space Astronomy Center (ESAC), P.O. Box 78,
     E-28691 Villanueva de la Ca\~nada, Madrid, Spain 
\and Instituut voor
     Sterrenkunde, KU Leuven, Celestijnenlaan 200B, 3001 Leuven,
     Belgium 
\and
     Department of Astrophysics, IMAPP, Radboud University Nijmegen, POBox 9010,
     6500 GL Nijmegen, the Netherlands }

\date{}
\authorrunning{Sarro et al.}
\titlerunning{Comparative clustering analysis of variable stars}

 
  \abstract
{Discovery of new variability classes in large surveys using
  multivariate statistics techniques such as clustering, relies
  heavily on the correct understanding of the distribution of known
  classes as point processes in parameter space.}
{Our objective is to analyze the correspondence between the classical
  stellar variability types and the clusters found in the distribution
  of light curve parameters and colour indices of stars in the CoRoT
  exoplanet sample. The final aim is to help in the identification on
  new types of variability by first identifying the well known
  variables in the CoRoT sample.}
{ We apply unsupervised classification algorithms to identify clusters
  of variable stars from modes of the probability density
  distribution. We use reference variability databases (Hipparcos and
  OGLE) as a framework to calibrate the clustering
  methodology. Furthermore, we use the results from supervised
  classification methods to interpret the resulting clusters.}
{We interpret the clusters in the Hipparcos and OGLE LMC databases in
  terms of large-amplitude radial pulsators in the classical
  instability strip and of various types of eclipsing binaries. The
  Hipparcos data also provide clear distributions for low-amplitude
  nonradial pulsators.  We show that the preselection of targets for
  the CoRoT exoplanet programme results in a completely different
  probability density landscape than the OGLE data, the interpretation
  of which involves mainly classes of low-amplitude variability in
  main-sequence stars. Our findings will be incorporated to improve
  the supervised classification used in the CoRoT catalogue
  production, once the existence of new classes or subtypes will be
  confirmed from complementary spectroscopic observations.}
{}
\maketitle
%
%
\section{Introduction}

In the past decade, a plethora of new light curves of celestial objects has
become available to the astronomical community. The OGLE project in its second
stage (OGLE II), e.g., provided astronomers with a total number of {\sl I}-band
magnitude time series of the order of 40 million. Ongoing 
variability-focused projects
such as ASAS \citep{asas}, MOST \citep{most} or CoRoT \citep{corotbook}, 
together with other future survey databases such as those
expected from the Panoramic Survey Telescope \& Rapid Response System
(abbreviated as 
Pan-STARRS\footnote{\tt http://pan-starrs.ifa.hawaii.edu/public/}), 
the Large Synoptic Survey Telescope 
(LSST\footnote{\tt http://www.lsst.org/lsst}) 
or Gaia\footnote{\tt http://www.rssd.esa.int/Gaia}, 
will provide us with a unique oportunity to discover new
variability classes, either thanks to their unprecedented capability to detect
weak periodic signals (the MOST and CoRoT cases) or to the large number of
objects observed (Gaia, Pan-STARRS or the LSST).

The discovery of new types of objects in large astronomical databases is
facilitated through the application of statistical techniques. In particular, in
the regime of very few cases in a potentially new class, we would be dealing
with a problem of outlier detection, whereas a sufficient number of examples in
the new class allows for the use of clustering techniques for class
discovery. In both cases, {\sl a priori} knowledge of the spatial distribution
of the known classes is essential.

In the general case, a variability survey can be seen as a collection of
celestial objects for which we have determined a series of parameters describing
the variability in the time series. In this work we concentrate on periodic
signals of variable stars as a starting point for understanding the variability
diversity that can be found in any of the aforementioned surveys. Thus, typical
attributes used to describe the time series are the significant detected
frequencies, fourier decomposition characteristics (harmonic amplitudes,
amplitude ratios, phase differences, etc.) and colour indices. An example of the
list of attributes used for supervised classification purposes can be found in
\cite{paper2}. Each variable star can then be represented by a point in a
multidimensional space, the axes of which are the attributes mentioned above. We
can view the database as a realization of a random point process and we can
approximate its originating probability density function by means of parametric
or nonparametric density estimation methods. In the outlier detection regime,
exotic or unusual objects will thus be defined as occupying regions of the
parameter space characterized by low probability densities. This further
requires a deep understanding of the noise properties in the random process,
like spurious frequency detections or the errors for low amplitude signals. In
the clustering regime, it is hoped that a good clustering algorithm is capable
of detecting the new class as a cluster.

A few clustering studies of variable celestial objects have been performed in
the past. \cite{2005MNRAS.358...30E} studied ASAS variability data using
Autoclass \citep{autoclass}. They manually separated two types of variables
according to the regularity of the light curves. Some 45\% of the total sample
of 1731 stars was considered to have a sufficiently regular behaviour and was
classified using Fourier decomposition coefficients. The remaining 55\% was
characterized by their period and second, third and fourth moments of their
light curve distribution.  In the first group, Autoclass identified nine
clusters. These were linked with eclipsing binaries (three clusters, two of
which are relatively pure and a third one mixed with ambiguous light curves and
wrong periods), Cepheids (two clusters, one with clear candidates, the other
with dubious cases), RR~Lyrae pulsators (one cluster with only four stars),
Small Amplitude Variables (a generic name for a mixture of low signal-to-noise
variables), and a last group without interpretation. Judging by the plots in
\cite{2005MNRAS.358...30E}, the latter cluster is characterized by periods in
the range 0.1 to 40 days and low $R_{21}$ values, where $R_{21}$ is defined as
the ratio between the amplitude of the first harmonic and the amplitude of the
frequency itself, for the dominant frequency.  Moreover, the corresponding phase
difference of these two Fourier terms, $\phi_{12}$, has random values which
points to almost sinusoidal light curves.


\cite{2004MNRAS.353..369B} applied Self-Organized Maps (hereafter SOMs) to study
the clustering structure of ROTSE preclassified light curves. They also applied
the algorithm to an artificial dataset of light curves for calibration
purposes. They folded the time series with the period and binned the
observations in a fixed number of phase bins. The training of the SOM was
carried out on a set of light curves previously classified in one of the
following categories: detached eclipsing binaries, contact eclipsing binaries,
$\delta$~Scuti stars, RR~Lyrae ab,RR~Lyrae c, and Cepheids. The results indicate
that, with this approach and dataset, SOMs are capable of separating detached
binaries, contact binaries and RR~Lyrae ab stars in three distinct categories,
whereas RR~Lyrae c, $\delta$~Scuti stars and Cepheids are mixed in one
cluster. As the authors point out, the lack of information on the period is the
main cause for this confusion of otherwise easily separable classes. The SOM
approach has several major drawbacks. First of all, it does not incorporate
information known to be relevant for the separation of clusters (e.g., the
harmonics of detected frequencies). Also, the binning process renders the method
inapplicable for databases where incomplete phase coverage is common (e.g., the
Hipparcos and Gaia databases). Further, it is not robust against spurious
frequency detections, a common problem with eclipsing binaries where identical
eclipses result in a detected frequency which is twice the orbital one. The
folded light curve will thus not resemble the eclipsing binaries
prototypes. Finally, the results have been obtained with a dataset which is the
result of intensive selection, i.e., only high signal-to-noise variables with
well determined periods and clear class assignments were used in the training of
the SOM.

In this work we characterize the distribution of variable objects in parameter
space. Natural groups formed in several reference variability databases, such as
the Hipparcos, the OGLE II Large Magellanic Cloud, and the CoRoT exoplanet
programme are considered.  Any clustering analysis is intrinsically limited by
the sampling properties of the database to which it is applied. Thus, we can
only hope to describe the realm of stellar variability to the limits of each
experiment and the algorithms used to describe it, i.e., its limiting magnitude,
sampling properties, sensitivity and the efficiency of the frequency detection
algorithms. The capability of characterizing weak multiperiodic signals in faint
stars in the Magellanic Cloud is limited, e.g., and clustering results in that
regime cannot produce clean samples for this database.  In our approach, we keep
all variable stars in the database for which at least one statistically
significant frequency was detected, regardless whether they have been previously
classified in other works or not. Our sample thus is as realistic with respect
to the original probability density function as the data permits it to be.  The
goal is to understand as deeply as possible this point distribution, in order to
be prepared for class discovery in the presently considered and other
forthcoming variability databases resulting from Gaia, Pan-STARRS or LSST.

\section{Elements of a clustering approach}
\label{clustering}

There are many theoretical approaches to clustering in the specialized
statistics literature. In astronomy, the two most popular clustering techniques
are Bayesian parametric modelling (e.g., the Autoclass implementation) and
SOMs. There are several reviews describing unsupervised classification or
clustering techniques (see e.g. \cite{331504}). A description of recent advances
in the field is out of the scope of this article, but we believe that a brief
discussion of some of the main issues involved in this kind of multivariate
analysis is necessary for the subsequent understanding and interpretation of our
results.

Clustering techniques can be put into several categories depending on
the viewpoint used to characterize them. For example, they can be
divided into parametric and nonparametric techniques according to the
hypothesis underlying the method. Nonparametric techniques make no a
priori assumption regarding the shape of the dataset constituent
clusters. A good example of nonparametric techniques are density based
clustering techniques that identify cluster centres as the modes in a
kernel based estimate of the point density. Obvious drawbacks of this
approach are its incapacity to detect low contrast overdensities in
the vicinity of dominant clusters if they do not result in local
maxima, and the limitations inherent to the bias-variance trade-off
dependence of the kernel widths. Other nonparametric techniques
include hierarchical clustering and Nearest Neighbours related
algorithms. 

On the other hand, parametric techniques are developed under
the hypothesis that the database is a sample from a probability
density distribution that either is a functional form (model) of the
parameters, or can be approximated with a number of such
functions. The simplest (yet useful) model is to assume that the
data are generated from a distribution that is a linear combination of
multivariate normal distributions with unit covariance matrices. This
results in spherical clusters and linear boundaries between
clusters. This simple model has the mean of the distribution and the
width of the multivariate normal as parameters, and a popular
algorithm to find them is the Expectation-Maximization algorithm that
maximizes the likelihood of the data. The model can be made more
flexible and sophisticated by allowing for more complex covariance
matrices. This translates into clusters of unequal major axis in
arbitrary orientations (i.e., not necessarily aligned with the
parameter space axis). This is the case of Autoclass, except for the
fact that it goes beyond maximum likelihood estimation by introducing
priors for the parameters, thus turning into a fully Bayesian
inference. 

One advantage of parametric models is that they permit the detection of clusters
not associated with local maxima, such as those missed by density based
methods. Unfortunately, the evident disadvantage is that, if the underlying
assumption is incorrect or insufficient (i.e., if the clusters are not samples
from the hypothesized parametric model), the resulting clustering is
uninterpretable. This drawback is not present in non parametric clustering,
where clusters of arbitrary shape can be detected. Obviously, it is always
possible to recontruct the real shape of a cluster by adding more and more model
components (multivariate Gaussians in the most popular choice), but then the
possibility to detect new stellar populations is compromised because the
clusters are no longer linked to differences in the point samples, but are
created by the inability of the algorithm to fit the real distribution of
points.

Two other elements, central to the problem of finding a good clustering
description of a dataset, are the automatic determination of an optimal number
of clusters, and that of feature selection. The former is closely related to the
clustering evaluation measures, since choosing a number of clusters relies on
the availability of an objective measure of clustering quality. In the framework
of parametric clustering, the classical tools for model selection are readily
available. In the Bayesian approach, the Bayes factor is commonly used, which
incorporates a natural Occam's razor by penalizing solutions with excessive
number of parameters (clusters). Obviously, one can find a perfect match to any
density distribution by adding more and more components to our mixture
model. This would be the equivalent to the concept of overfitting in supervised
classification or regression problems. Other ways to assess clustering quality
related to maximum likelihood estimators are the Bayesian Information Criterion
(BIC) and Akaike Information Criterion (AIC), or the Median Split Silouhettes
(MSS, \cite{Pollard02amethod}) and the gap statistic
(\cite{TibsPollard02amethodhirani00estimatingthe}), which can also be applied to
nonparametric clustering.


The other critical aspect of clustering experiments is the selection
of an optimal attribute set, capable of incorporating the attributes
where clustering occurs but, at the same time, discarding those
irrelevant to treat the problem. Unfortunately, this field has not evolved
as much in unsupervised as in supervised classification, and there is
no commonly accepted technique to perform attribute selection. In the
experiments described in Sect.\,\ref{experiments}, we 
selected the attributes manually based on the past experience in supervised
classification. 

One last aspect that has to be considered in the selection of an
algorithm for the clustering analysis of variablity data is that of
scalability. Both the time and space complexities of the algorithms,
and the possibility of parallelizing the data processing has to be
considered, specially if the algorithms have to be applied to
large databases such as those expected in the next few years.

This summary is not intended to be exhaustive but rather brings an introduction
to the key aspects of the problem of class discovery in variability
databases. There is simply no best clustering methodology for all types of
problems. We stress the importance of making both the assumptions underlying
each method and the evaluation criteria explicit before a clustering experiment
is designed.

\section{\label{method}Adopted methodology}

In designing the experiments and methodology described below, we imposed several
requirements. The solution proposed here is an attempt to fulfil all of the
following requirements:

\begin{enumerate}

\item {\sl Interpretability}. We consider that no class discovery is
  possible if the clusters obtained by a given method are not
  interpretable in terms of the experts domain knowledge. In this
  case, the domain is that of regular variable stars and this means
  that the clusters defined by the method have to recover the
  classical variability types reasonably well. This requirement
  obviously has to be related to the parameter space where clustering
  is performed: it can not be hoped to separate variability classes
  only differing in some spectroscopic signature (e.g., due to
  metallicity effects) if the parameter space is built only with the
  properties of a photometric time series.

\item The algorithms involved in the clustering process have to be
scalable to the orders of magnitude typical of the era of astronomical
surveys. Nowadays, large scale variability surveys like those expected to
come from Gaia, Pan-STARRS or LSST, will be producing parameters for
$\approx 10^9$ objects or more, and class discovery techniques have to be
capable to deal with these enormous databases.

\item The methodology has to be model free or, at least, use the most
flexible models available, in order to avoid that the clusters obtained
are dictated by the model selected. 

\end{enumerate}

In the following, we describe the clustering methodology adopted in
the performed experiments. We decided to use the so-called density
based approach to clustering, whereby clusters are identified with
maxima of the density distribution, and points are assigned to their
closest maximum or mode. This density is never actually computed and
points are assigned to modes (clusters) by using the Modal EM
(Expectation Maximization, MEM) algorithm \citep{HMAC}, which can be
summarized as follows. Let $S=\{x_1, x_2,\ldots,x_n\}$ be the set of
objects $x_i\in{\cal R}^d$ to be clustered. The $d$ components in each
$x_i$ are the attributes or parameters used to characterize the
variable objects (frequencies, amplitudes, phase differences, colour
indices, etc.; see below). Let $f(x)=\sum_{k=1}^K f_k(x)$ be the
density of points approximated by a Mixture Model with $K$
components. In our case, we used multivariate normal distributions
${\cal N}_{\mu, \Sigma}(x)$, with mean $\mu$ and covariance matrix
$\Sigma$. Then, given a point $x$, the MEM algorithm consists in
alternating the following two steps, starting with $r=0$ and
$x^{(0)}=x$:

\begin{enumerate}

\item {\sl Expectation:} compute the posterior probability of each
mixture component, $p_k=\frac{\pi_k f_k(x^{(r)})}{f(x^{(r)})}$;

\item {\sl Maximization:} update $x^{(r+1)} = argmax_x \sum_{k=1}^{K}p_k
\log(f_k(x))$.

\end{enumerate}

When applied to one object in the database defined by $x_i$, the MEM
algorithm converges to a local maximum of the density
estimate. Clustering is then accomplished by grouping points that
converged to the same mode (to within some tolerance). Our starting
point is a kernel density estimate such that, if the database contains
$n$ objects, $f(x)$ is defined by a mixture model with $n$ components,
one associated to each object:

\begin{equation}
f(x) = \sum_{k=1}^{n} \frac{1}{n} {\cal N}_{x_k,\Sigma_k}(x).
\end{equation}  

Given that the implementation has to be capable of processing datasets
with a number of instances of the order of several times $10^4$ (in
these experiments) and larger, the implementation makes use of
$k$-dimensional trees ($kd$--trees hereafter; see e.g. \cite{355745} and
\cite{358850}).

It is evident that the MEM algorithm involves an approximation to the
density which, in this case, is a kernel based
approximation. Therefore, depending on the kernel bandwidth $\Sigma_k$
used in the estimate, we will have different sets of modes: less nodes
and large clusters for a large kernel width, and many small clusters
when the kernel bandwidth is small. At each scale we are analysing
different aspects of the dataset but, as the kernel width is made
smaller, the modes of the density distribution can be made to
correspond to local maxima due to noise in the finite sampling of the
point process. Therefore, the smallest scales one is ready to accept
have to be judged on physical terms in order to establish the
significance of the clusters obtained.

In our case, we carried out the analysis in a hierarchical manner, starting with
large kernel bandwidths to define the large groupings in the data. Thereafter,
we decreased the widths until we could no longer establish the meaningfulness of
the largest clusters obtained. We also implemented adaptive kernel widths in
order to avoid large variances in the clustering results in the low density
regions. The kernel width adaptation takes into account the local density of
points in order to increase the width in regions where the nearest neighbours
are at large distances and thus, fixed kernel estimation would result in a local
maximum of the density at each observed object.

Since these experiments were designed to serve as the basis for subsequent class
discovery, we concentrated the description of the clusters in terms of known
classical variables. Thus, the clustering decomposition (the decrease in the
kernel bandwidths) stopped when we no longer could interpret the resulting
clusters.

\section{Performed experiments}
\label{experiments}

In the following, we describe the experiments we performed with three databases:
the Hipparcos variability archive \citep{HIPPARCOS}, the OGLE database of Large
Magellanic Cloud variables \citep{Zebrun:2001}, and the CoRoT database
corresponding to the first four runs (IRa01, SRc01, LRc01, and LRa01;
\cite{paper3}). These databases are characterized by very different statistical
properties. While the sampling is quite regular but drastically different 
in the OGLE and CoRoT databases,
this is less so in the Hipparcos case. The number of cases in each sample is
also very different: 43351 variabels were retrieved from the OGLE LMC database,
2419 variables from Hipparcos and 14642 from CoRoT (see below for selection
criteria). Finally, colours are available for very large fractions of the OGLE
and Hipparcos databases whereas the CoRoT sample is analysed only based on
parameters derived from the time series.

The figures given above for the number of objects in each database are the
result of selecting sources with some predefined criteria. For both the OGLE and
Hipparcos databases of variable stars, we selected objects with available $V-I$
colour index. For the Hipparcos and CoRoT databases we also requested $p$-values
in the frequency detection phase below $10^{-7}$ (see \cite{paper3} for a
definition of the $p$-values). 

The original time series in each database was processed in order to detect
statistically significant frequencies. The procedure was described in detail in
\cite{paper1}, \cite{paper2}, and \cite{paper3} and is therefore omitted here.
We only retain five relevant attributes in the analysis below, namely the first
detected frequency (denoted as $\nu$ in the OGLE and Hipparcos cases and as
$\nu_1$ in the CoRoT case and expressed in d$^{-1}$ throughout the paper), 
the amplitude of the first component in the Fourier
decomposition of this frequency $A_{11}$, the $R_{21}$ ratio which stands for
the ratio of the amplitudes of the first two components in the Fourier
decomposition of the first frequency, the phase difference $\phi_{12}$ between
the second and first harmonic components in the Fourier decomposition and (in
the OGLE and Hipparcos cases), the $V-I$ colour index. In the CoRoT case, given
the quality of the time series, we used the amplitudes of the first four
components in the Fourier decomposition of the first detected frequency, and the
first two of the second detected frequency $\nu_2$.  Further attributes can be
added in order to look deeper into the composition of the clusters found. This
includes parameters related to higher orders in the Fourier decomposition and
additional undereddened colour indices if available, such as $J-H$, or
$H-K$. The latter only make sense for selected clusters where multiperiodicity
or infrared emission are attributes expected to result in clustering structures
at smaller scales.

In the following, we present and describe clustering results obtained
with the density based methodology described in section
\ref{method}. As explained above, clusterings with different levels of
detail can be obtained with different kernel bandwidths $\Sigma_k$
(the covariance matrix used in the normal distributions placed at each
point). Thus, starting with a large kernel bandwidth (resulting in few
large clusters), it is possible to refine and subdivide the clusters
obtained by using increasingly smaller bandwidths. The resulting
structure resembles a tree (the dendrogram) where clusters (branches)
subdivide into smaller clusters at smaller scales.

The covariance matrix $\Sigma_k$ can be chosen isotropic
($\Sigma_k=\sigma^2\cdot \mathbbm{1}$) with $\sigma^2$ being the tuning
parameter effectively controlling the kernel bandwidth), or, more
generally, anisotropic. In this work we have made the diagonal
elements proportional to the nearest neighbour distance. The
proportionality is introduced by normalizing the covariance matrix
such that the maximum along the diagonal is made equal to the tuning
parameter $\sigma^2$. 

The parameters selected for the clustering process and the level at
which the dendrogram is cut are mostly dictated by the
interpretability of the clusters and clarity of plots. While it is
possible to present purer clusters (i.e., with more homogeneous
components) it is usually at the expense of a larger number of
clusters. It is important to remark that each dimension of the
  data was normalized in the range $[0,1]$ to avoid attributes with
  larger absolute values to dominate the kernel variance. Therefore,
  the input data to the clustering algorithm are contained inside a
  unit hypercube. This means that the quoted values of $\sigma$ (the
  kernel bandwidth) are relative to this unit length defining the
  range in each dimension. The values used for each experiment
  (Hipparcos, OGLE LMC and CoRoT) were 0.2, 0.15 and 0.2
  respectively. While, as explained above, the availability of colour
  indices and class labels in the first two experiments guided us in
  the selection of the actual value of the kernel width, in the case
  of the CoRoT database it was based on the visual inspection of
  density contour and scatter plots. The clumpiness of the
  distribution of data points in parameter space was somewhat less
  than for any of the other two experiments, suggesting a larger value
  of $\sigma$. We nevertheless decided to maintain the same value
  $\sigma=0.2$, given the unprecedented quality of the CoRoT data.

Plots of the clusters will be numbered and referred to according to
the cluster identifier.

\subsection{The Hipparcos archive}

\begin{figure*}[htp]
   \centering
   \includegraphics[scale=.50]{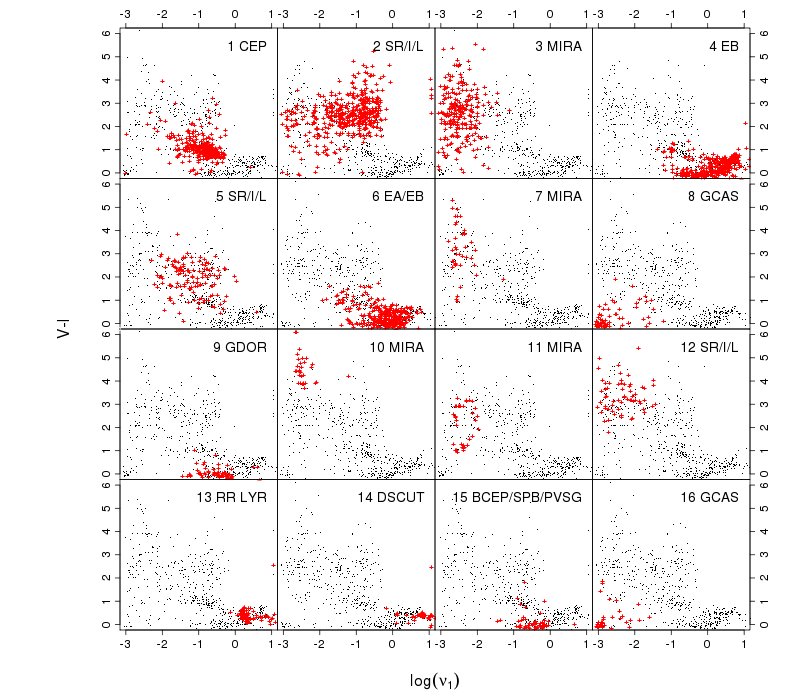}
   \caption{The clustering structure of the Hipparcos variable stars
     archive at $\sigma=0.2$. The $x$-axis represents the logarithm of
     the frequency and the $y$-axis the $V-I$ colour index. Black
       dots represent the complete database and red crosses identify
       cluster members. Predominant classes according to the Hipparcos
       variability annex are included as labels in each box.}
   \label{hip-f1-vmi}
\end{figure*}

\begin{figure*}[H]
   \centering
   \includegraphics[scale=0.5]{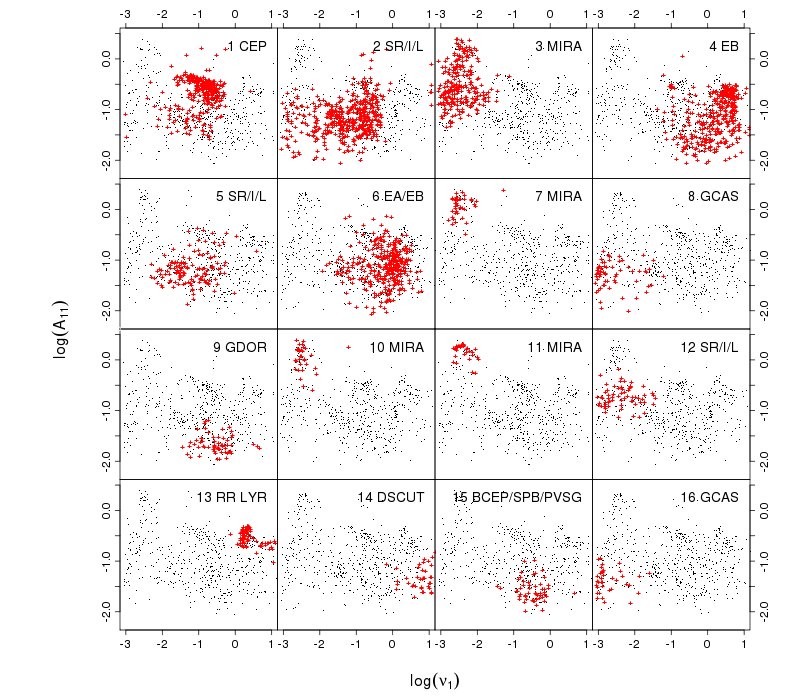}
   \caption{The clustering structure of the Hipparcos archive at
     $\sigma=0.2$. The $x$-axis represents the logarithm of the
     frequency and the $y$-axis, the logarithm of the amplitude of the
     first component in the Fourier decomposition. Black dots
       represent the complete database and red crosses identify
       cluster members.}
   \label{hip-f1-A11}
\end{figure*}

\begin{figure*}[H]
   \centering
   \includegraphics[scale=0.5]{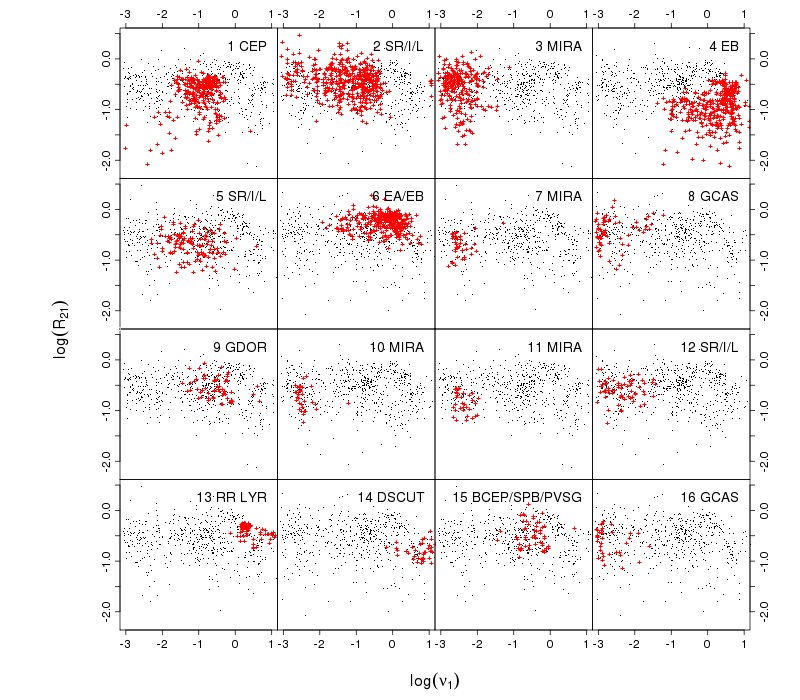}
   \caption{The clustering structure of the Hipparcos archive at
     $\sigma=0.2$. The $x$-axis represents the logarithm of the
     frequency and the $y$-axis, the logarithm of the $R_{21}$ ratio
     between the amplitudes of the first two components in the Fourier
     decomposition. Black dots represent the complete database and
       red crosses identify cluster members.}
   \label{hip-f1-R21}
\end{figure*}

\begin{figure*}[H]
   \centering
   \includegraphics[scale=0.5]{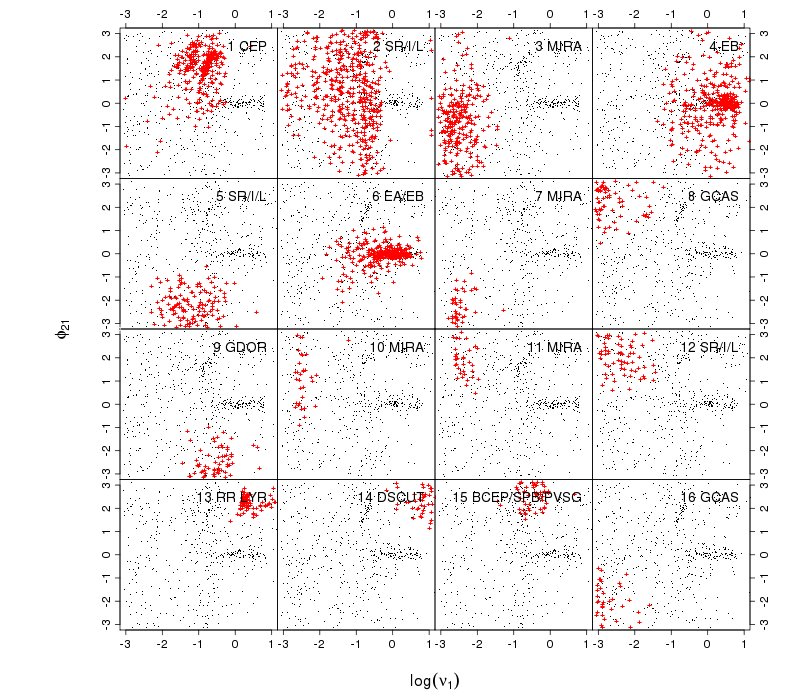}
   \caption{The clustering structure of the Hipparcos archive at
     $\sigma=0.2$. The $x$-axis represents the logarithm of the
     frequency and the $y$-axis, the phase difference $\phi_{12}$
     between the first two components in the Fourier
     decomposition. Black dots represent the complete database and
       red crosses identify cluster members.}
   \label{hip-f1-fi21}
\end{figure*}

\begin{figure*}[H]
   \centering
   \includegraphics[scale=0.5]{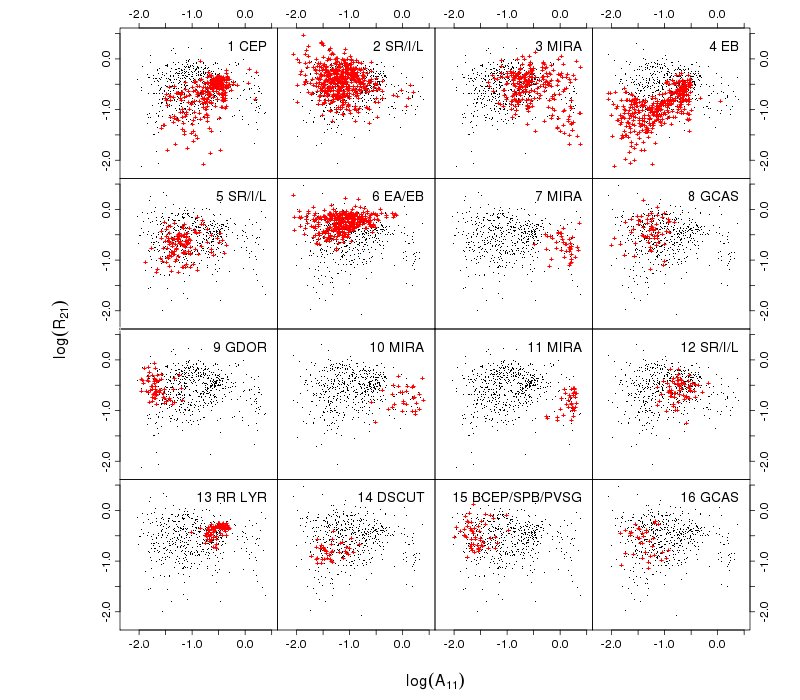}
   \caption{The clustering structure of the Hipparcos archive at
     $\sigma=0.2$. The $x$-axis represents the logarithm of the
     amplitude of the first component in the Fourier decomposition,
     and the $y$-axis, the logarithm of the $R_{21}$ ratio between the
     amplitudes of the first two components. Black dots represent
       the complete database and red crosses identify cluster
       members.}
   \label{hip-A11-R21}
\end{figure*}

\begin{figure*}[H]
   \centering
   \includegraphics[scale=0.5]{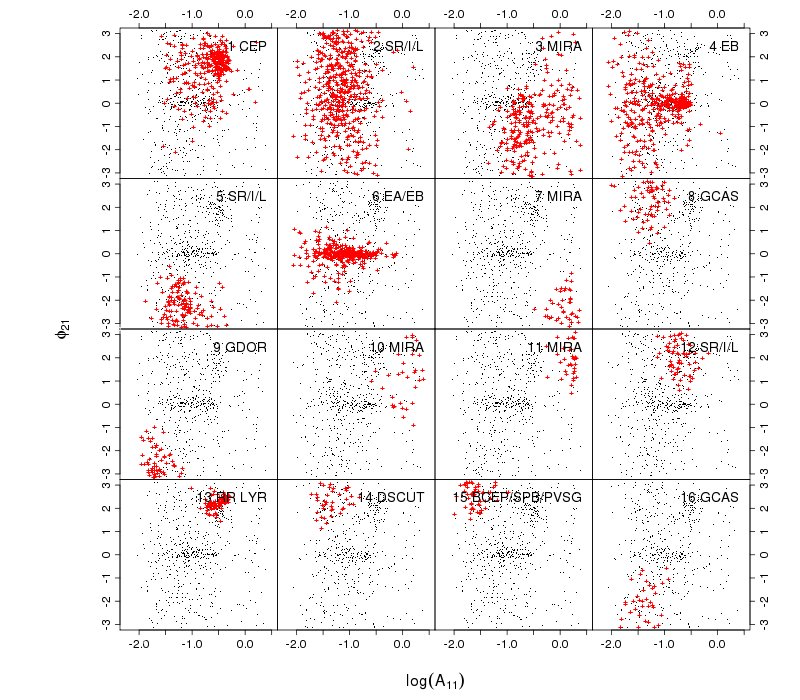}
   \caption{The clustering structure of the Hipparcos archive at
     $\sigma=0.2$. The $x$-axis represents the logarithm of the
     amplitude of the first component in the Fourier decomposition,
     and the $y$-axis, the phase difference $\phi_{12}$ between the
     first two components in the Fourier decomposition. Black dots
       represent the complete database and red crosses identify
       cluster members.}
   \label{hip-A11-fi21}
\end{figure*}

\begin{figure*}[H]
   \centering
   \includegraphics[scale=0.5]{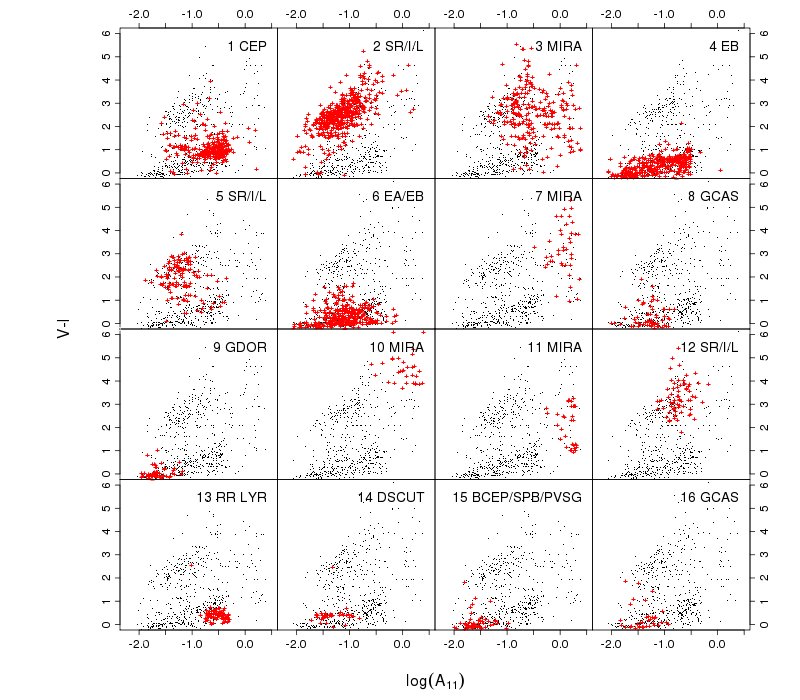}
   \caption{The clustering structure of the Hipparcos archive at
     $\sigma=0.2$. The $x$-axis represents the logarithm of the
     amplitude of the first component in the Fourier decomposition,
     and the $y$-axis, the $V-I$ colour index. Black dots represent
       the complete database and red crosses identify cluster
       members.}
   \label{hip-A11-VMI}
\end{figure*}

\begin{figure*}[H]
   \centering
   \includegraphics[scale=0.5]{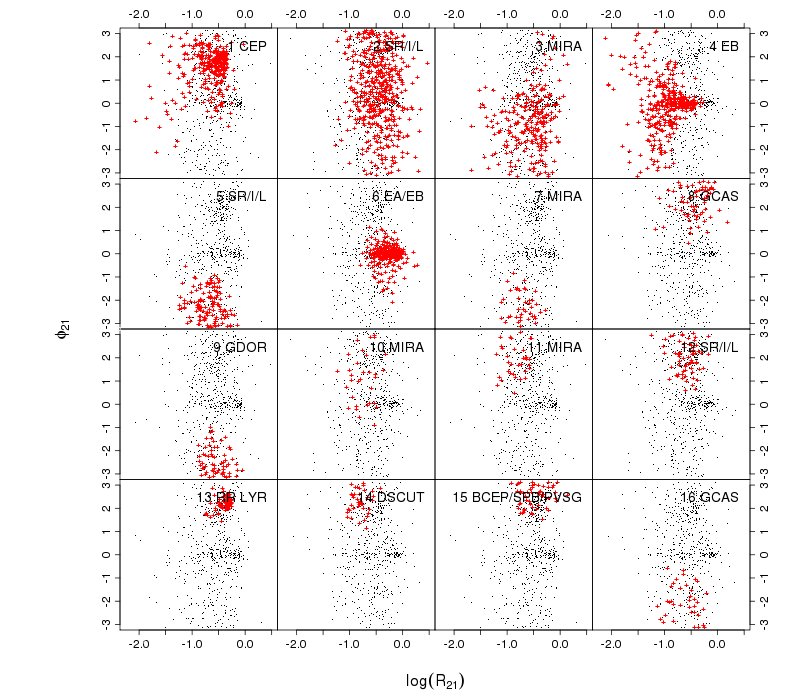}
   \caption{The clustering structure of the Hipparcos archive at
     $\sigma=0.2$. The $x$-axis represents the logarithm of the
     $R_{21}$ ratio between the amplitudes of the first two components
     of the Fourier decomposition, and the $y$-axis, the phase
     difference $\phi_{12}$ between the first two components in the
     Fourier decomposition. Black dots represent the complete
       database and red crosses identify cluster members.}
   \label{hip-R21-fi21}
\end{figure*}

\begin{figure*}[H]
   \centering
   \includegraphics[scale=0.5]{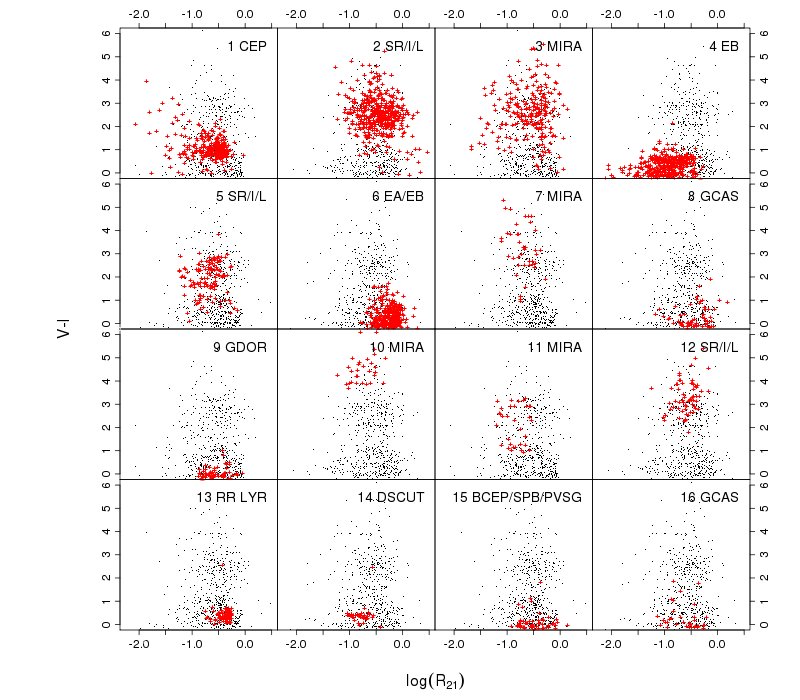}
   \caption{The clustering structure of the Hipparcos archive at
     $\sigma=0.2$. The $x$-axis represents the logarithm of the
     $R_{21}$ ratio between the amplitudes of the first two components
     of the Fourier decomposition, and the $y$-axis, the $V-I$ colour
     index. Black dots represent the complete database and red
       crosses identify cluster members.}
   \label{hip-R21-VMI}
\end{figure*}

\begin{figure*}[H]
   \centering
   \includegraphics[scale=0.5]{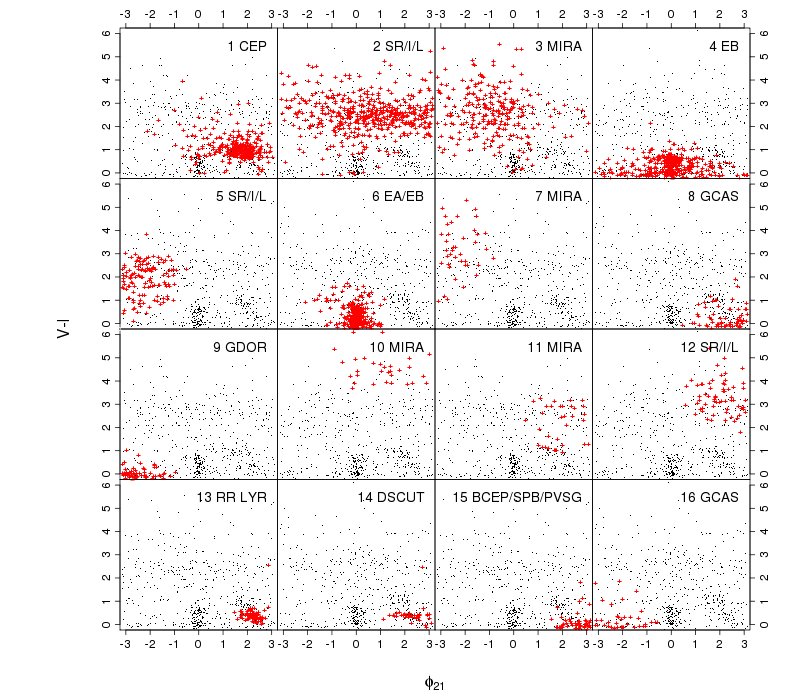}
   \caption{The clustering structure of the Hipparcos archive at
     $\sigma=0.2$. The $x$-axis represents the phase difference
     $\phi_{12}$ between the first two components in the Fourier
     decomposition, and the $y$-axis, the $V-I$ colour index.
       Black dots represent the complete database and red crosses
       identify cluster members.}
   \label{hip-fi21-VMI}
\end{figure*}

The entire Hipparcos variability catalogue was processed as described
in \cite{paper1}, and light curves with $p$-values, as defined in
\citet{paper3}, below $10^7$ were retained for clustering
analysis. Furthermore, we made use of class labels in order to
interpret the resulting clusters. Figure\,\ref{hip-f1-vmi} represents
the 16 clusters found in this set of objects in the
$\log(\nu)$--$(V-I)$ plane, using a value of the kernel width
$\sigma=0.2$.  Other projections of the original database using the
other attributes can be found in
Figures\,\ref{hip-f1-A11}-\ref{hip-fi21-VMI}.  For the discussion of the
Hipparcos database, we make use of the class labels from the General
Catalogue of Variable Stars \citep[GCVS]{gcvs} because this was also
done by the Hipparcos team \citep[e.g.]{eyergrenon}.

From Fig.\,\ref{hip-f1-vmi} we deduce that the algorithm separated long period
(i.e., low frequency) variables of similar properties but with different
behaviour in the values of $\phi_{12}$. Examples are clusters 7 and 10 which
share similar properties and are only separated in the value of the $\phi_{12}$
of their modes. This is obviously an artefact due to the use of the quantity
$\phi_{12}$ as an attribute, while this quantity is invariant with respect to
$2\pi$.  The phase difference $\phi_{12}$ is only useful for a few classes, like
RR~Lyrae pulsators, Cepheids or eclipsing binaries because they show highly
non-sinusoidal light curves. For long period variables (classes M, SR, I and L
in the nomenclature of the GCVS) or low-amplitude pulsators, we find a more
homogeneous coverage of the range of values of $\phi_{12}$, in agreement with
the physics of their oscillations and their nearly sinusoidal variations in the
light curve which implies a large uncertainty in the $R_{21}$ and
$\phi_{12}$ values.

Further, the algorithm succeeds in the recovery of the most conspicuous groups
of variables: cluster 1 contains most of the Cepheids in the Hipparcos
catalogue, cluster 13 most of the RR~Lyrae stars, cluster 4 the binaries of the
EB type, and cluster 6 binaries of the EA and EB types. The Long Period
Variables group (LPVs) includes as most numerous subgroups the Miras (M),
Semiregular Variables (SR), Irregular Variables (I) and Slow Irregular Variables
(L). These subtypes are spread over clusters 2, 5 and 12 (SR, I, and L), and 3,
7, 10, and 11 (Mira Variables with negligible contributions from other
LPVs). While cluster 12 is characterized by periods of the order of several
hundred days, clusters 2 and 5 show a much larger spread of periods down to a
few days. Cluster 3 contains the lower-amplitude Miras, whose $\phi_{12}$-values
are spread out compared to these for the Miras in clusters 7, 10, and 11. The
latter are artifically divided due to the $2\pi$ degeneracy in $\phi_{12}$ on
the one hand and due to redder colour indices on the other hand.

We interpret the separation of Miras into four distinct clusters as the result
of a finite sample, producing local low contrast maxima at random locations. A
larger kernel density would smooth out these small overdensities at the expense
of merging neighbouring clusters.  The obvious solution to this is to cut the
clustering dendrogram at different levels in each branch. Furthermore, this
allows for the selection of relevant attributes in each branch, the necessity of
which is exemplified by the segregation of groups introduced by an attribute
such as $\phi_{12}$ which is not relevant for all classes of variables but very
discriminating for others.

Cluster 14 contains most of the $\delta$ Sct stars in the sample while all other
contributions to this cluster are small. The second most numerous contributing
class are the SX\,Phe stars with only three objects. This is fully
understandable from a physical point of view, as the SX\,Phe stars' variability
behaviour is very similar to the high-amplitude $\delta$\,Sct stars but they are
old Population~II stars instead \citep{rodriguez}. Most other SX\,Phe stars
constitute a satellite group of the RR~Lyrae stars in cluster 13 (these two
groups separate for smaller values of the kernel width) and are not assigned to
the $\delta$ Sct cluster because they have larger amplitudes than the
prototypical Population~I $\delta\,$Sct stars and only somewhat smaller than the
Population~II RR\,Lyrae stars (see Fig.\,\ref{hip-f1-A11}). It is very assuring
that the algorithm manages to distribute the SX\,Phe stars over the two other
clusters whose physical properties they share.

Cluster 15 is composed of rather strictly periodic blue stars. These
are pulsating stars of spectral type B belonging to the $\beta\,$Cep,
slowly pulsating B (SPB) and periodically variable supergiant (PVSG)
classes discovered by means of multivariate discriminant analysis by
\citet{waelkens} and confirmed from ground-based follow-up data by
\citet{aerts00}, \citet{aerts99}, and \citet{lefever}.  These stars
are in general nonradial multiperiodic oscillators with low-order
pressure ($\beta\,$Cep) or gravity (SPB, PVSG) modes excited by the
$\kappa\,$mechanism.  Their amplitudes are not too different from
those of the $\delta\,$Sct stars, even though some of the class
members show somewhat more outspoken small nonlinear effects in their
oscillations. Their periods are typically an order of magnitude longer
than those of the $\delta\,$Sct stars, as nicely illustrated in
Fig.\,\ref{hip-f1-vmi}. The ground-based follow-up studies have shown
that some of these candidate pulsating B stars turned out to be
ellipsoidal binaries \citep{decat00} or spotted stars
\citep{briquet04} with periods similar to the oscillation periods of
SPBs, but this was the case for a small minority only.

Cluster 9, on the other hand, contains $\gamma\,$Dor stars. These are
F-type multiperiodic stars pulsating in nonradial gravity modes with
periods similar to those of the SPBs, i.e., also an order of magnitude
longer than the $\delta\,$Sct stars. They are situated on the red side
of the classical instability strip. Numerous new class members were
indeed found in the Hipparcos database by \citet{aerts98} and
\citet{handler}. Most of these were confirmed as class member later
on, from extensive follow-up studies \citep{aerts04,mathias04,decat06}
just as for the B-type pulsators discussed above.  The amplitude and
phase behaviour of the $\gamma\,$Dor stars is similar to the one of
the $\delta\,$Sct, $\beta\,$Cep and SPB stars, taking into account the
2$\pi$ degeneracy for $\phi_{12}$.

Finally, clusters 8 and 16 are dominated by so-called $\gamma$
Cassiopeiae stars (GCAS in the GCVS nomenclature). The star
$\gamma\,$Cas is a member of the classical Be stars, which are objects
showing Balmer line emission in their spectrum due to the presence of
a circumstellar disc (see \citet{gcas} for an extensive review on the
observational and physical properties of those stars).  This
inhomogeneous class consists of both single and binary stars and,
moreover, some Be stars show oscillations and/or light curve trends
while others do not. It is thus normal that these objects are spread
out over various clusters.  Clusters 8 and 16 contain 32 and 17 such
stars, respectively. The difference between the two cluster modes is
only due to the phase difference $\phi_{12}$ of $2\pi$, which is again
artificial. Cluster 15 described above, also contains 9 Be stars
according to the Hipparcos classification.

We conclude that our clustering analysis turned out to be a powerful method to
separate the monoperiodic large-amplitude classical variables and binaries from
the low-amplitude multiperiodic nonradial pulsators. Since these populations are
well represented in the Hipparcos database of variables, we have shown our
methodology to have the potential to recover all these classes in unexplored new
databases as well, by means of the attributes that we have used here.  Moreover,
our analysis divided the four classes of nonradial pulsators along the main
sequence, i.e., the best targets for asteroseismology (see \citet{aerts08} for a
review), in four different clusters with the appropriate properties according to
the physics of these stars.

\subsection{The OGLE archive}

The OGLE LMC variability database was processed in the same way as the Hipparcos
archive and an equivalent dataset was generated with the same
attributes. However, the phase difference $\phi_{12}$ was not used in the
clustering experiments. While the use of $\phi_{12}$ was convenient to avoid a
strong contamination of the RR~Lyrae and Cepheids clusters by a particular type
of eclipsing binaries for the Hipparcos database, the denser sampling (43351
objects) defines the maxima of the density distribution better in the OGLE LMC
case and diminishes (but does not eliminate) the contamination. Rather, the
Wessenheit index $W_I\equiv I-1.55(V-I)$, which is a brightness indicator
independent of extinction, was added in the analysis since this quantity has
been used previously in the mining of the OGLE database.

\begin{figure*}[htp]
   \centering
   \includegraphics[scale=.50]{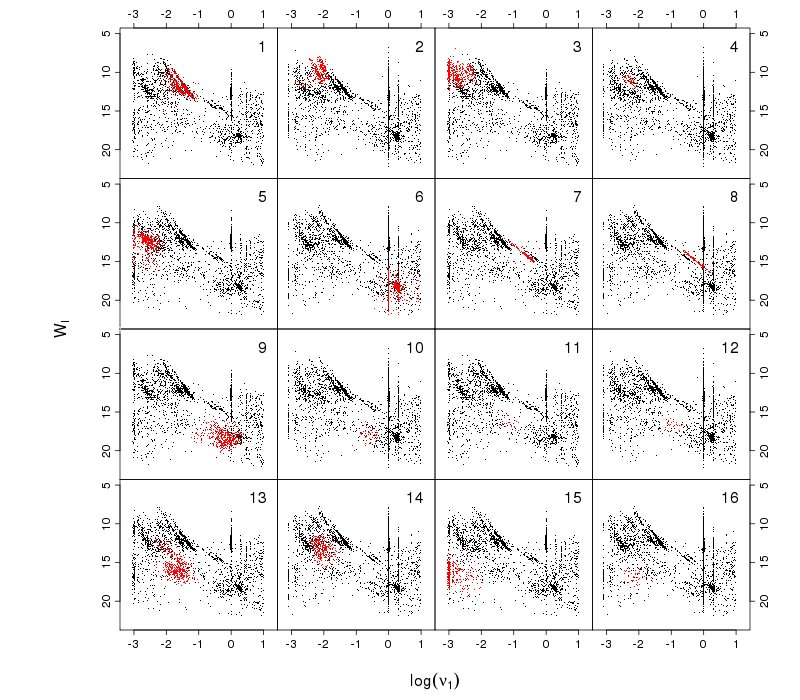}
   \caption{The clustering structure of the OGLE LMC archive at
     $\sigma=0.15$. The $x$-axis represents the logarithm of the
     frequency and the $y$-axis, the Wessenheit index $W_I$.
       Black dots represent the complete database and red crosses
       identify cluster members.} Note the occurrence of spurious
     frequencies at a value of one per day and multiples thereof due
     to the daily gaps in the data (one-day aliasing problem).
   \label{lmc-f1-wi}
\end{figure*}

\begin{figure}[htp]
   \centering
   \includegraphics[scale=.30]{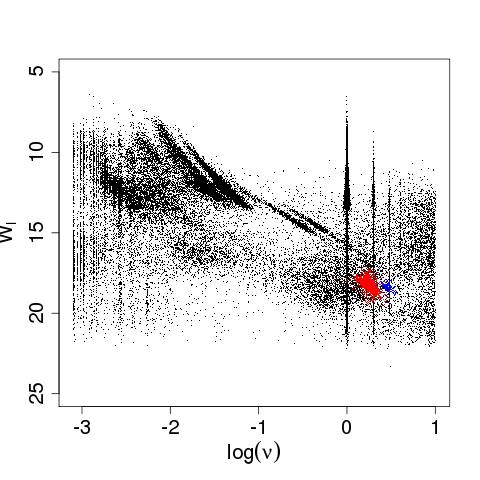}
   \caption{The two largest subgroups in the RR~Lyrae cluster,
     obtained with a smaller kernel bandwidth ($\sigma=0.1$) than in
     Fig.\,\protect\ref{lmc-f1-wi}. They correspond to the well known
     loci of the RRab (red dots) and RRc/RRd (blue dots) subtypes. The
     $x$-axis represents the logarithm of the frequency and the
     $y$-axis, the Wessenheit index $W_I$.}
   \label{lmcrrlyr}
\end{figure}

\begin{figure*}[H]
   \centering
   \includegraphics[scale=0.5]{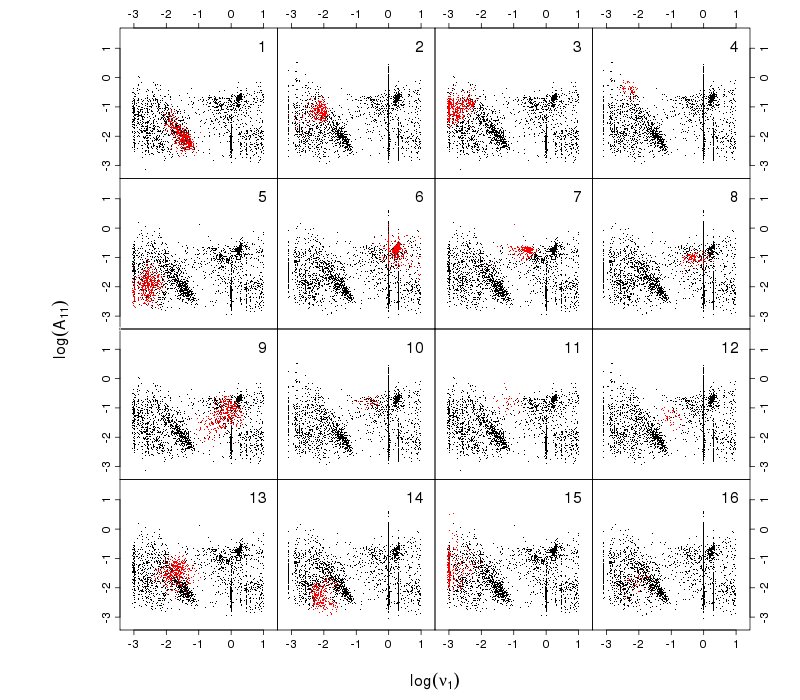}
   \caption{The clustering structure of the OGLE LMC archive at
     $\sigma=0.15$. The $x$-axis represents the logarithm of the
     frequency and the $y$-axis, the logarithm of the amplitude of the
     first component in the Fourier decomposition.
       Black dots represent the complete database and red crosses
       identify cluster members.}
   \label{lmc-f1-A11}
\end{figure*}

\begin{figure*}[H]
   \centering
   \includegraphics[scale=0.5]{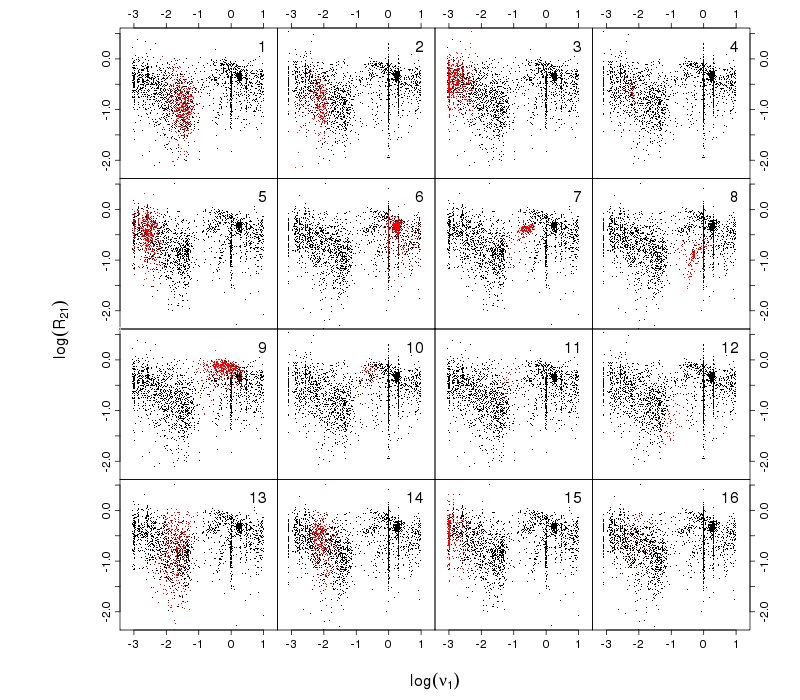}
   \caption{The clustering structure of the OGLE LMC archive at
     $\sigma=0.15$. The $x$-axis represents the logarithm of the
     frequency and the $y$-axis, the logarithm of the $R_{21}$ ratio
     between the amplitudes of the first two components in the Fourier
     decomposition. Black dots represent the complete database
       and red crosses identify cluster members.}
   \label{lmc-f1-R21}
\end{figure*}

\begin{figure*}[H]
   \centering
   \includegraphics[scale=0.5]{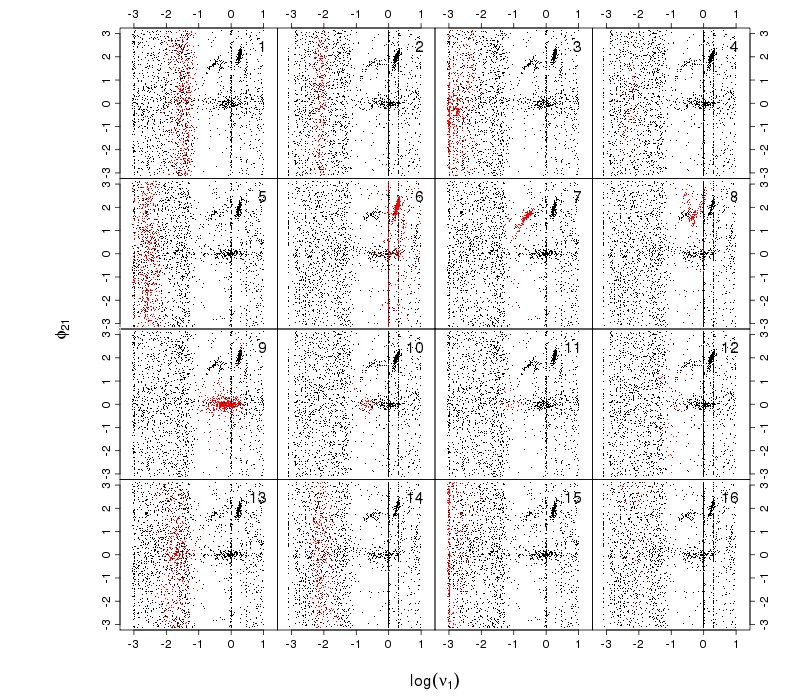}
   \caption{The clustering structure of the OGLE LMC archive at
     $\sigma=0.15$. The $x$-axis represents the logarithm of the
     frequency and the $y$-axis, the phase difference $\phi_{12}$
     between the first two components in the Fourier
     decomposition. Black dots represent the complete database
       and red crosses identify cluster members.}
   \label{lmc-f1-fi21}
\end{figure*}

\begin{figure*}[H]
   \centering
   \includegraphics[scale=0.5]{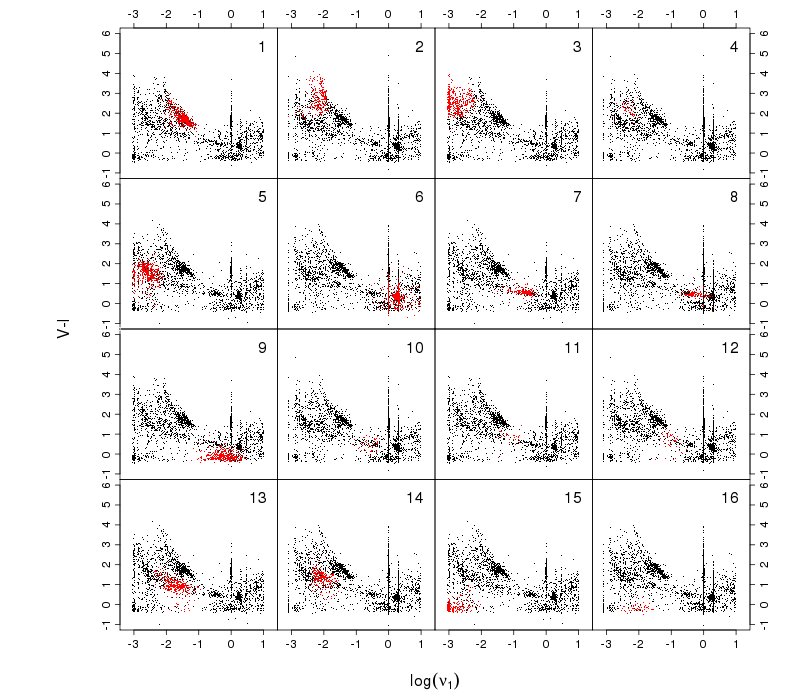}
   \caption{The clustering structure of the OGLE LMC archive at
     $\sigma=0.15$. The $x$-axis represents the logarithm of the
     frequency and the $y$-axis, the $V-I$ colour index. Black
       dots represent the complete database and red crosses identify
       cluster members.}
   \label{lmc-f1-vmi}
\end{figure*}

\begin{figure*}[H]
   \centering
   \includegraphics[scale=0.5]{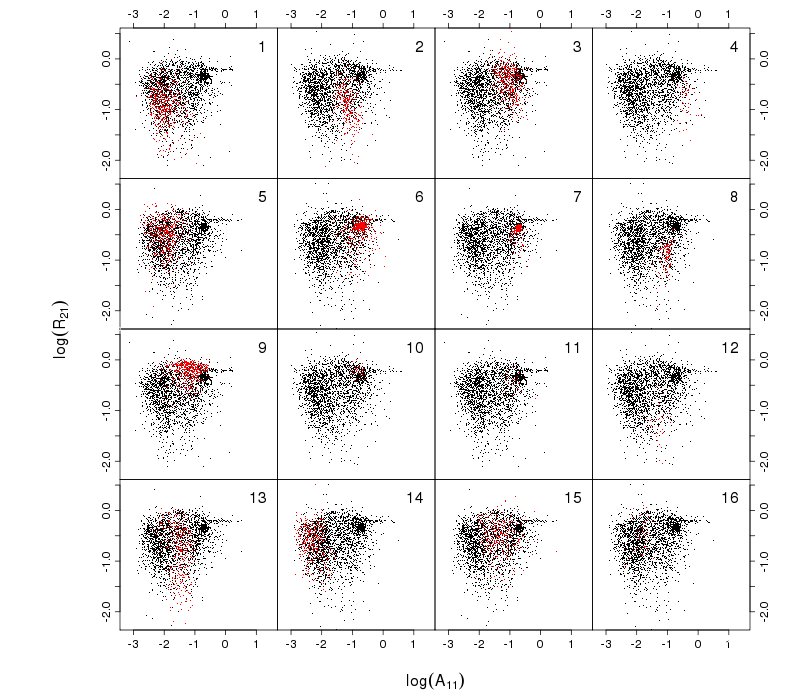}
   \caption{The clustering structure of the OGLE LMC archive at
     $\sigma=0.15$. The $x$-axis represents the logarithm of the
     amplitude of the first component in the Fourier decomposition,
     and the $y$-axis, the logarithm of the $R_{21}$ ratio between the
     amplitudes of the first two components. Black dots represent
       the complete database and red crosses identify cluster
       members.}
   \label{lmc-A11-R21}
\end{figure*}

\begin{figure*}[H]
   \centering
   \includegraphics[scale=0.5]{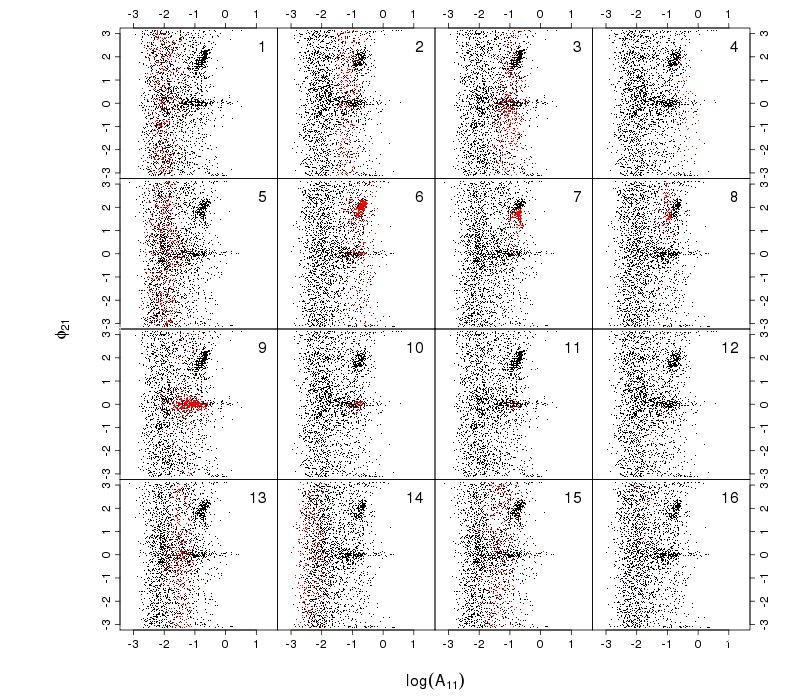}
   \caption{The clustering structure of the OGLE LMC archive at
     $\sigma=0.15$. The $x$-axis represents the logarithm of the
     amplitude of the first component in the Fourier decomposition,
     and the $y$-axis, the phase difference $\phi_{12}$ between the
     first two components in the Fourier decomposition. Black
       dots represent the complete database and red crosses identify
       cluster members.}
   \label{lmc-A11-fi21}
\end{figure*}

\begin{figure*}[H]
   \centering
   \includegraphics[scale=0.5]{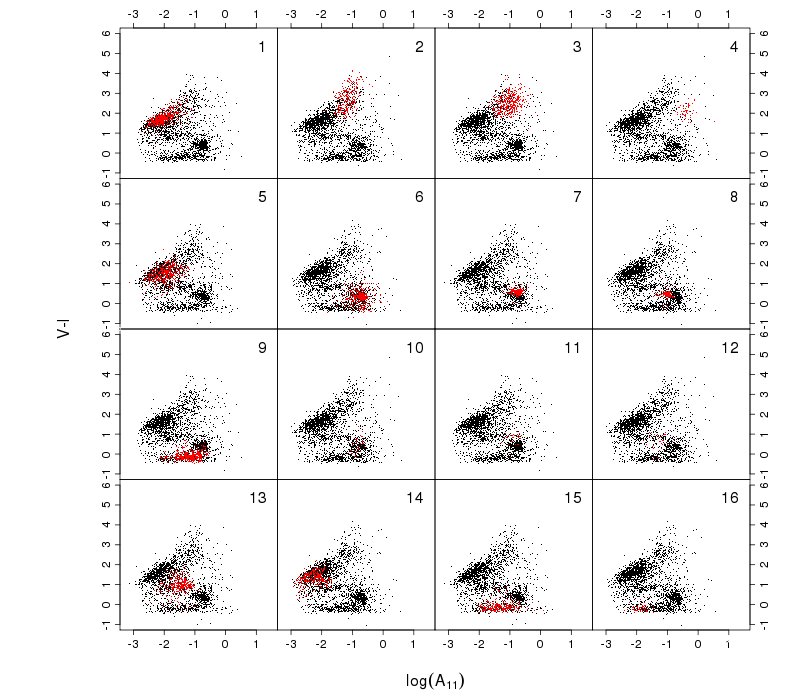}
   \caption{The clustering structure of the OGLE LMC archive at
     $\sigma=0.15$. The $x$-axis represents the logarithm of the
     amplitude of the first component in the Fourier decomposition,
     and the $y$-axis, the $V-I$ colour index. Black dots
       represent the complete database and red crosses identify
       cluster members.}
   \label{lmc-A11-VMI}
\end{figure*}

\begin{figure*}[H]
   \centering
   \includegraphics[scale=0.5]{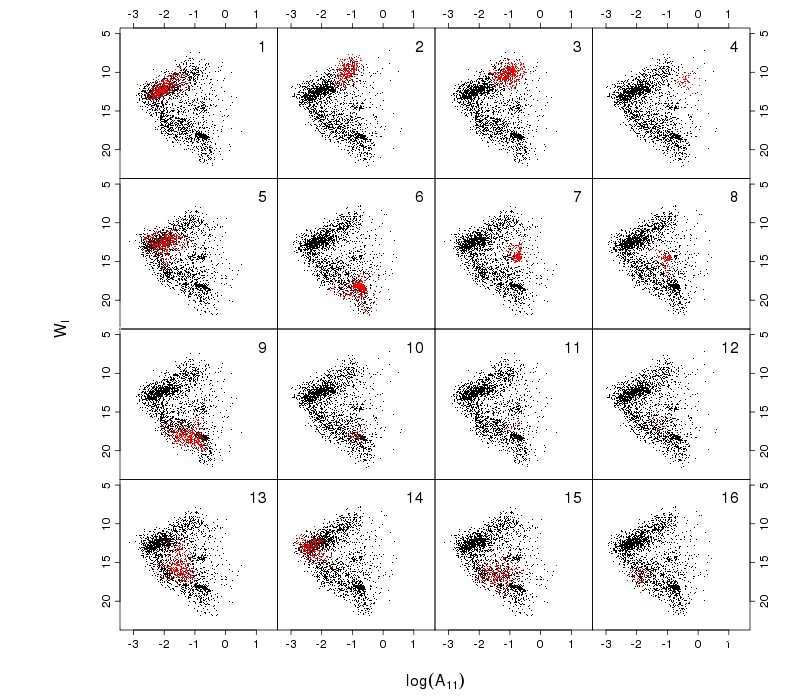}
   \caption{The clustering structure of the OGLE LMC archive at
     $\sigma=0.15$. The $x$-axis represents the logarithm of the
     amplitude of the first component in the Fourier decomposition,
     and the $y$-axis, the Wessenheit index $W_I$. Black dots
       represent the complete database and red crosses identify
       cluster members.}
   \label{lmc-A11-wi}
\end{figure*}

\begin{figure*}[H]
   \centering
   \includegraphics[scale=0.5]{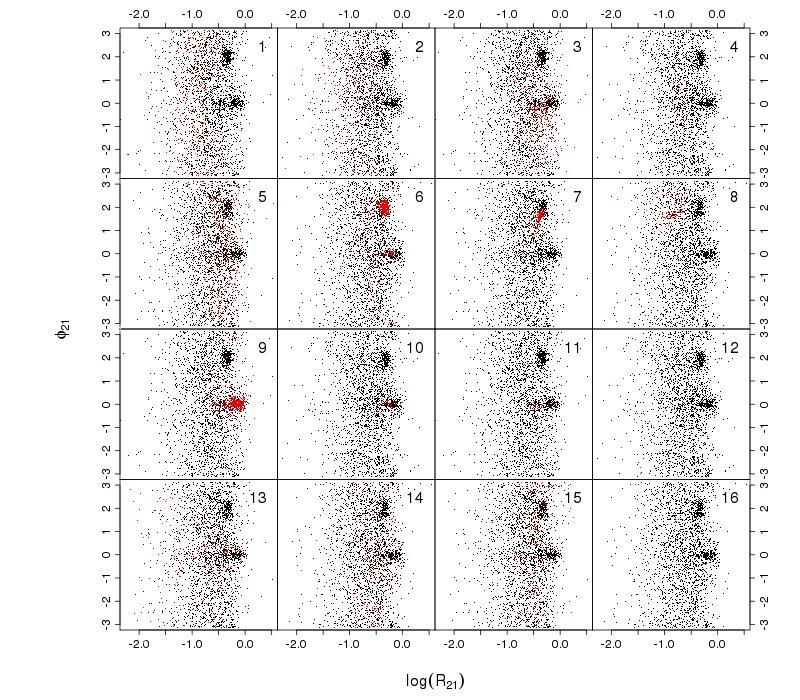}
   \caption{The clustering structure of the OGLE LMC archive at
     $\sigma=0.15$. The $x$-axis represents the logarithm of the
     $R_{21}$ ratio between the amplitudes of the first two components
     of the Fourier decomposition, and the $y$-axis, the phase
     difference $\phi_{12}$ between the first two components in the
     Fourier decomposition. Black dots represent the complete
       database and red crosses identify cluster members.}
   \label{lmc-R21-fi21}
\end{figure*}

\begin{figure*}[H]
   \centering
   \includegraphics[scale=0.5]{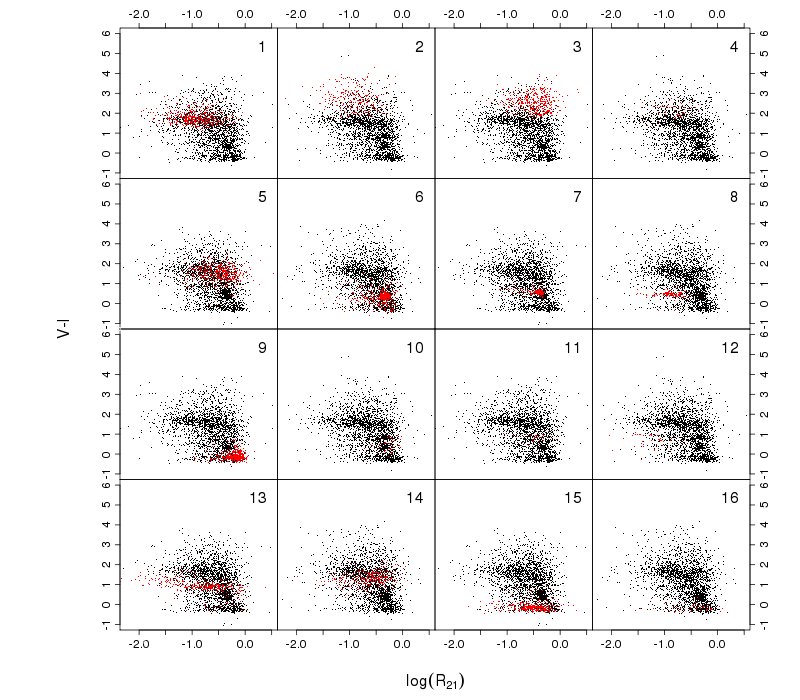}
   \caption{The clustering structure of the OGLE LMC archive at
     $\sigma=0.15$. The $x$-axis represents the logarithm of the
     $R_{21}$ ratio between the amplitudes of the first two components
     of the Fourier decomposition, and the $y$-axis, the $V-I$ colour
     index. Black dots represent the complete database and red
       crosses identify cluster members.}
   \label{lmc-R21-VMI}
\end{figure*}

\begin{figure*}[H]
   \centering
   \includegraphics[scale=0.5]{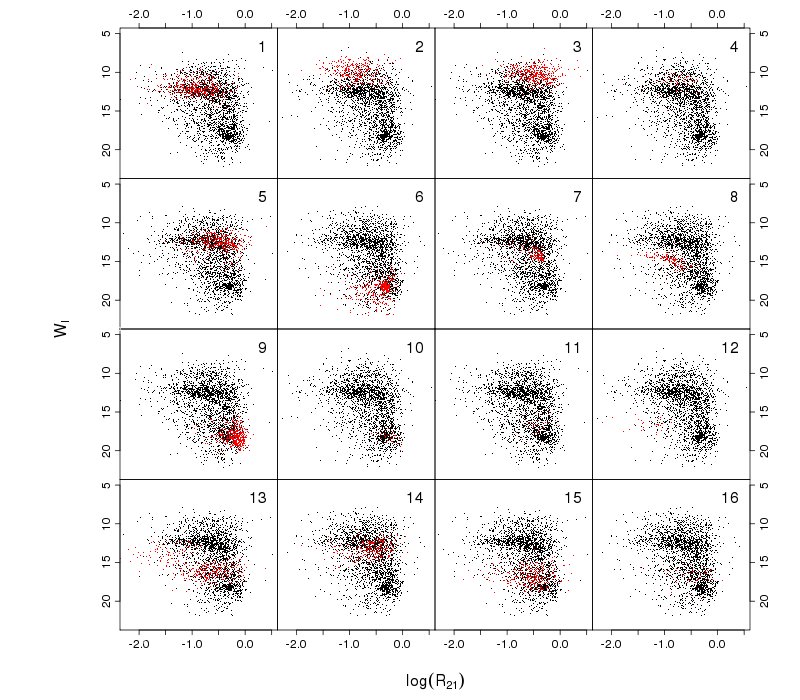}
   \caption{The clustering structure of the OGLE LMC archive at
     $\sigma=0.15$. The $x$-axis represents the logarithm of the
     $R_{21}$ ratio between the amplitudes of the first two components
     of the Fourier decomposition, and the $y$-axis, the Wessenheit
     index $W_I$. Black dots represent the complete database and
       red crosses identify cluster members.}
   \label{lmc-R21-wi}
\end{figure*}

\begin{figure*}[H]
   \centering
   \includegraphics[scale=0.5]{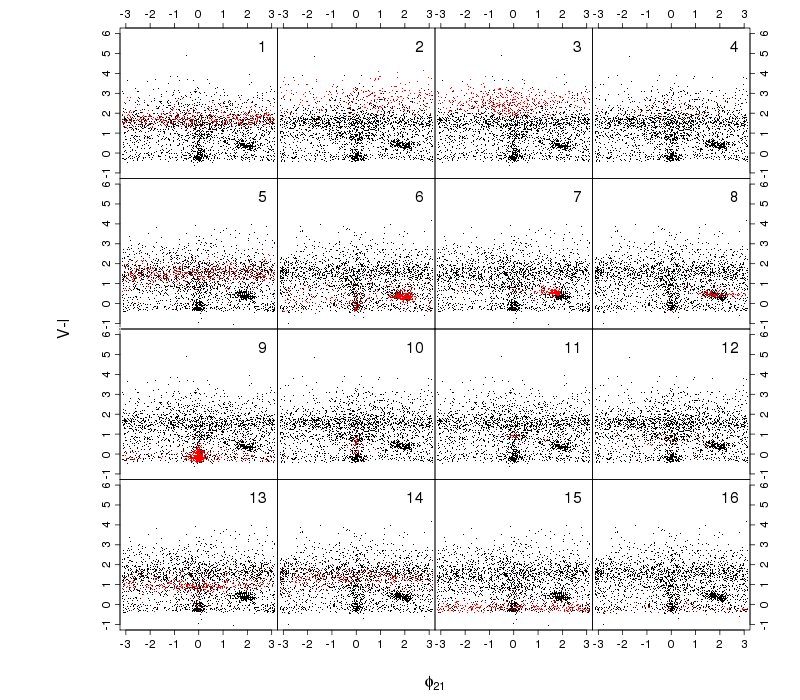}
   \caption{The clustering structure of the OGLE LMC archive at
     $\sigma=0.15$. The $x$-axis represents the phase difference
     $\phi_{12}$ between the first two components in the Fourier
     decomposition, and the $y$-axis, the $V-I$ colour index. 
       Black dots represent the complete database and red crosses
       identify cluster members.}
   \label{lmc-fi21-VMI}
\end{figure*}

\begin{figure*}[H]
   \centering
   \includegraphics[scale=0.5]{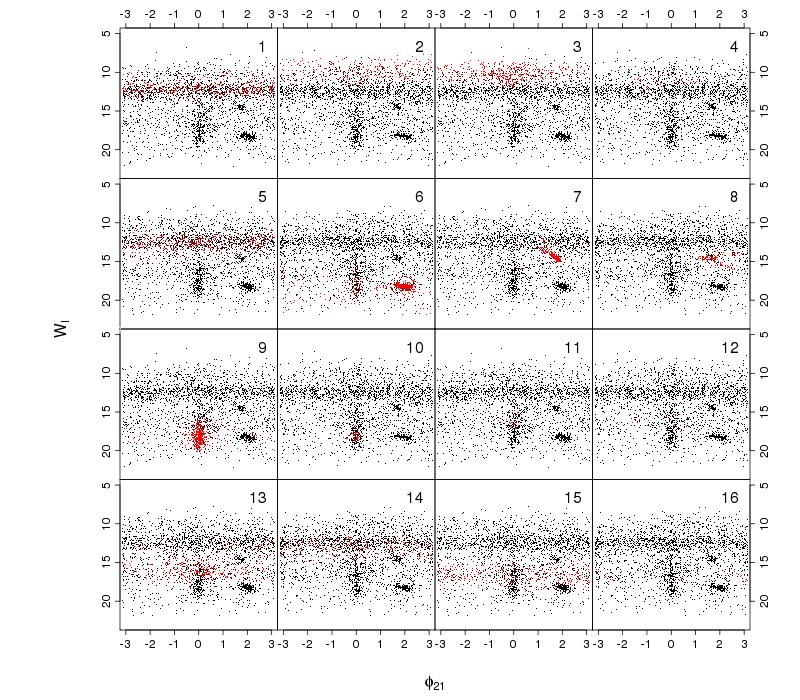}
   \caption{The clustering structure of the OGLE LMC archive at
     $\sigma=0.15$. The $x$-axis represents the phase difference
     $\phi_{12}$ between the first two components in the Fourier
     decomposition, and the $y$-axis, the Wessenheit index $W_I$. 
       Black dots represent the complete database and red crosses
       identify cluster members.}
   \label{lmc-fi21-wi}
\end{figure*}

\begin{figure*}[H]
   \centering
   \includegraphics[scale=0.5]{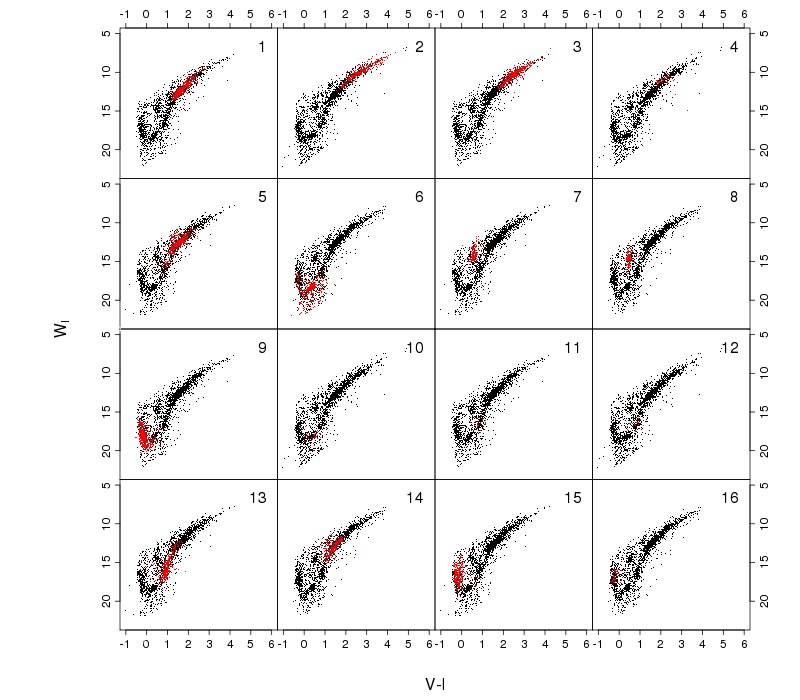}
   \caption{The clustering structure of the OGLE LMC archive at
     $\sigma=0.15$. The $x$-axis represents the $V-I$ colour index,
     and the $y$-axis, the Wessenheit index $W_I$. Black dots
       represent the complete database and red crosses identify
       cluster members.}
   \label{lmc-vmi-wi}
\end{figure*}

The list of well identified clusters at the largest scales is similar to the one
described above for the Hipparcos database, except that the OGLE data are less
suited to find low-amplitude multiperiodic variables. This is illustrated in
Fig.\,\ref{lmc-f1-wi} in the frequency-$W_I$ index plane, and further clarified
in other projections in Figs\,\ref{lmc-f1-A11}--\ref{lmc-vmi-wi}. Amongst the
largest clusters are those containing 
the First Overtone (FO) and Fundamental Mode (FU) Cepheids
(clusters 7 and 8, respectively), and the RR~Lyrae (cluster 6). 
The RR~Lyrae cluster is further separated at smaller scales, into two large
clusters corresponding i) to the ab subtype, and ii) to a mixture of the c
subtype and double mode RR~Lyrae pulsators (see Fig.\,\ref{lmcrrlyr}), and
several smaller clusters containing mainly spurious detections.

\begin{figure}[htp]
   \centering
   \includegraphics[scale=.30]{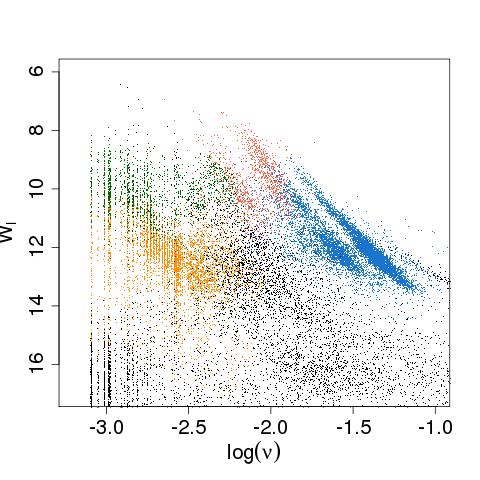}
   \caption{The clustering structure of the OGLE LMC archive at
     $\sigma=0.15$, in the Long Period Variables region. The $x$-axis
     represents the logarithm of the frequency and the $y$-axis, the
     Wessenheit index $W_I$. Blue dots represent the A and B sequences
     (cluster 1), red dots represent the C' sequence (cluster 2), dark
     green corresponds to the C sequence and the upper
     part of the D sequence (cluster 3) and orange represents the D
     sequence (cluster 4).}
   \label{lmclpv1}
\end{figure}

The Long Period Variables sequences defined by \cite{Soszynski:2005} in the
$W_{JK}$-$\log(P)$ plane (where the $W_{JK}$ index is the equivalent of the
Wessenheit index for the $J$ and $K_S$ near infrared photometric bands) are
grouped into four large categories. The Long Period Variables region in
Fig.\,\ref{lmc-f1-wi} is shown again for clarity in Fig.\,\ref{lmclpv1}. Each of
these clusters is further separated into smaller groups at finer scale levels in
the clustering hierarchy. In order of decreasing frequencies (increasing
periods), the first and largest cluster comprises the A and B sequences (the
so-called OGLE Small Amplitude Red Giant Stars, OSARGS); the adjacent cluster
corresponds to the C' sequence (cluster 2; Mira and Semiregular variables) in
the $W_{JK}$-$\log(P)$ plane. Fig.\,\ref{lmclpv1} shows that this
cluster can be separated into several groups at smaller bandwidth scales. The
reason for this is that the unique C' sequence in the $W_{JK}$-$\log(P)$ plane,
splits into two when the Wessenheit index $W_I$ is used instead of $W_{JK}$,
with each ridge characterized by a different slope corresponding to the
different chemical composition of the stars (O-rich in the case of the larger
frequency cluster and C-rich in the other cluster). The C sequence in the
$W_{JK}$-$\log(P)$ plane is barely visible when the Wessenheit index is used,
but the algorithm detects a cluster (cluster 3) that coincides with the
descriptions found in \cite{Soszynski:2005}, except for the fact that it also
contains objects in the D sequence (those with redder $V-I$ colours and larger
amplitudes). Cluster 4 also seems to correspond to objects in the C
sequence. Finally, the D sequence is clearly visible at the smallest
frequencies. As mentioned above, these groups (obtained with a large kernel
bandwidth) are further split into the well known Long Period Variables sequences
in subsequent levels of the hierarchy obtained with smaller kernel bandwidths.

\begin{figure*}[htp]
   \centering
   \includegraphics[scale=.5]{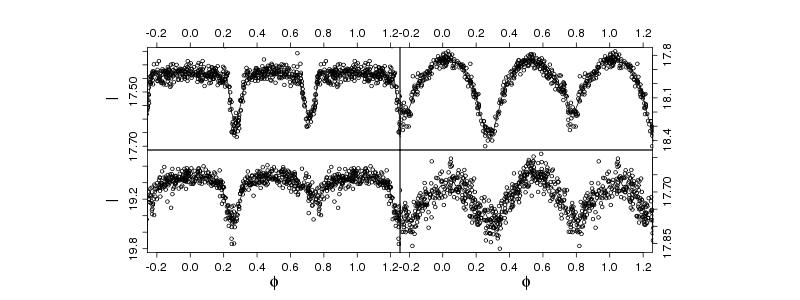}
   \caption{Example light curves from objects in clusters 9 (top
     left), 10 (bottom left), 11 (top right), and 12 (bottom right),
     of the OGLE LMC sample.}
   \label{lmc-ex-ecls}
\end{figure*}

The eclipsing binaries and ellipsoidal variables are gathered in 5 clusters
(clusters 9 to 13). Cluster 9 corresponds to detached binaries of the EA and EB
types, with flat light curves between eclipses. As we go from cluster 9 to
cluster 12, we see a sequence in the shapes of the eclipsing binaries light
curves, in the sense that the eclipse widths increase until there no longer
exists a flat part of the light curve between
eclipses. Figure\,\ref{lmc-ex-ecls} shows several examples of the light curves
of objects closest to the corresponding cluster modes.
Following the results obtained by \citet{paper2} regarding the classification of
hot main-sequence pulsators, some of these objects may have ended up in these
binary star clusters.

\begin{figure}[htp]
   \centering
   \includegraphics[scale=.30]{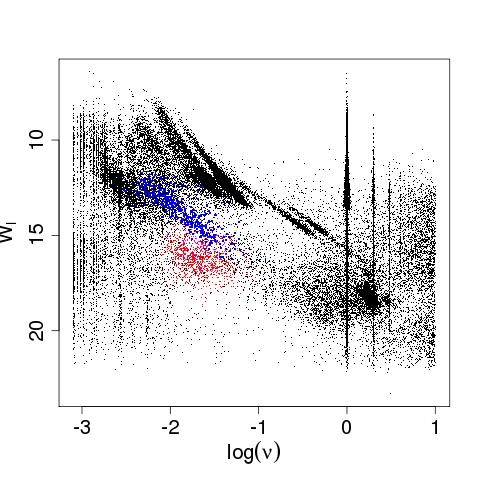}
   \caption{The clustering structure of the OGLE LMC archive at
     $\sigma=0.3$ (after renormalization to the unit hypercube)
     for cluster 10. The $x$-axis represents the logarithm of the
     frequency and the $y$-axis, the Wessenheit index $W_I$. It
     clearly shows how the cluster is divided into two subgroups
     corresponding to the locus of ellipsoidal variables (blue)
     and EW eclipsing binaries (red).}
   \label{sub7}
\end{figure}

\begin{figure}[htp]
   \centering
   \includegraphics[scale=.30]{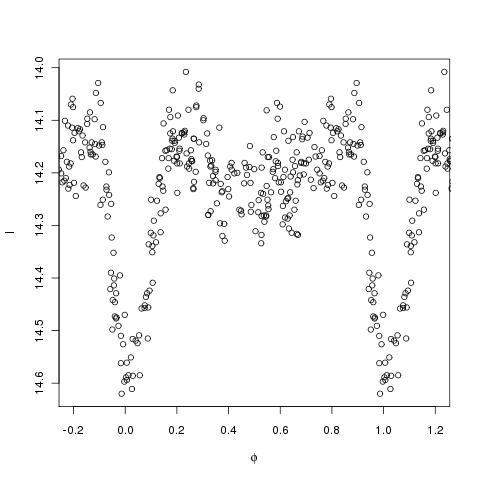}
   \caption{Example of an ellipsoidal variable in an eccentric system
     (cluster 13b). The light curve qualitatively resembles the
     synthetic light curve computed for an eccentricity value $e=0.2$
     and periastron length $p=90\deg$ shown in Fig.\,8 of
     \cite{2004AcA....54..347S}}
   \label{lmc-ex-ells}
\end{figure}

Cluster 13 contains two subclusters that are easily separated at smaller kernel
bandwidths (see Fig.\,\ref{sub7}). One of them (cluster 13a shown in blue in
Fig.\,\ref{sub7}) shows a clear correlation between the frequency and the
Wessenheit index, in agreement with the slope found by
\cite{2004AcA....54..347S}, while cluster 13b (shown in red in Fig.\,\ref{sub7})
shows no correlation in the $log(\nu)$-$W_I$ plane. Visual inspection of the
light curves in this cluster seems to indicate a predominant component of EW
systems, although we also find eccentric ellipsoidals such as the one shown in
Fig.\,\ref{lmc-ex-ells}. The comparison of clusters 13a and 14 with the
population of variables discussed by \cite{2004AcA....54..347S} seems to suggest
that these two clusters are mainly composed of ellipsoidal variables in binary
systems.

The last two clusters in Fig.\,\ref{lmc-f1-wi} correspond to systems occupying
the same region in the parameter space where the Hipparcos Be stars are found
and correspond with stars having long periods and blue colours in the LMC sample
(see Fig.\,\ref{lmc-f1-vmi}).  Visual inspection of the light curves of systems
in the neighbourhood of the cluster mode indicates the presence of nonperiodic
behaviour with long time scales in cluster 15 and a mixture of nonperiodic and
strictly periodic behaviour in cluster 16.

\begin{figure*}[htp]
   \centering
   \includegraphics[scale=.40]{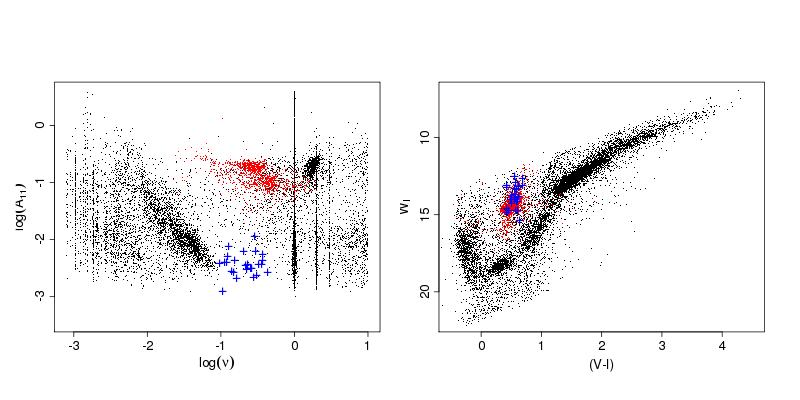}
   \caption{Low amplitude periodic variable candidates in two
     dimensional projections of the parameter space. In the left plot,
     the $x$ axis represents the logarithm of the frequency, and the
     $y$ axis the logarithm of the amplitude of the first harmonic
     component of the Fourier decomposition. In the right hand plot,
     the $x$ axis represents the $V-I$ colour index, and the $y$ axis
     the Wessenheit index $W_I$. In both plots, classical Cepheids
     pulsating in the fundamental mode or the first overtone are
     represented with red dots, and the low amplitude candidates with
     blue crosses.  }
   \label{laceps}
\end{figure*}

Additional clusters not shown in Fig.\,\ref{lmc-f1-wi} show
interesting properties, worth further investigation. One of them
(shown in Fig.\,\ref{laceps}) contains a selection of candidates to
the category of low amplitude periodic variables \citep{paper3}.  This
figure shows their position in two 2D projections of the parameter
space, together with the fundamental and first overtone Cepheids
loci. Two of the candidates (SC4 323401 and SC3 35239) belong to the
list of ultra low amplitude Cepheids in \cite{ULAS}. Other may again
represent some of the candidate slowly pulsating B stars found from
supervised classification in \citet{paper2}.

\begin{figure}[htp]
   \centering
   \includegraphics[scale=.30]{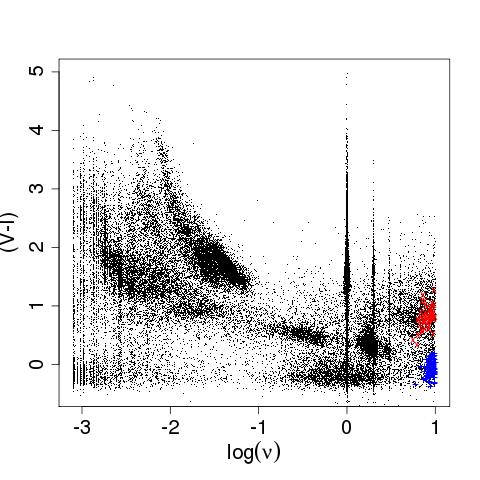}
   \caption{Blue plus signs and red crosses represent the small scale structure
     of one of the clusters found, not shown in
     Fig.\,\ref{lmc-f1-wi}. The $x$ axis corresponds to the logarithm of
     the frequency $\nu$ , and the $y$ axis, to the $V-I$ colour index.}
   \label{ws}
\end{figure}

Another cluster, shown in the $\log(\nu)$-$(V-I)$ plane in Fig.\,\ref{ws},
splits into several subclusters at smaller kernel bandwidths, two of which seem
to group stars with properties typical of the $\delta\,$Sct (shown as red dots)
and $\beta$ Cep (in blue) types, respectively. These correspond to some of the
stars identified as such type of main-sequence nonradial pulsators 
in \citet{paper3}. 

\begin{figure}[htp]
   \centering
   \includegraphics[scale=.30]{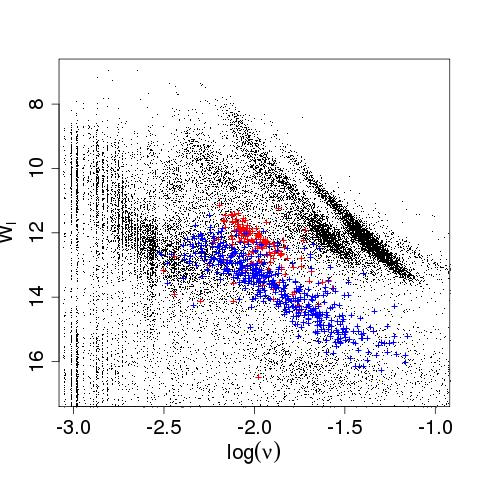}
   \caption{Red crosses represent stars in one of the clusters
     obtained, showing a strong correlation in the $\log(\nu)$-$W_I$
     plane. The $x$ axis corresponds to the logarithm of the frequency
     $\nu$ , and the $y$ axis, to the Wessenheit index $W_I$. The
     region shown corresponds to the red giant sequences, and blue
     crosses correspond to the cluster interpreted as ellipsoidal
     variables in the text (shown here for comparison).}
   \label{ws2}
\end{figure}

Finally, the cluster shown in Fig.\,\ref{ws2} is located in the region
occupied by the various sequences of red giant variables discussed in
\cite{Soszynski:2005} (see their Fig.\,3), but does not correspond to
any of the sequences discussed therein. It shows a remarkable
correlation in the $\log(\nu)$-$(V-I)$ plane that seems to continue
the sequence of C-rich red giants in the C sequence, although there
also exists the possibility that it represents a subpopulation of
ellipsoidal variables (shown in blue in Fig.\,\ref{ws2}).

We conclude to find reasonable agreement between the extractor-type and
supervised classification results applied to the OGLE LMC database and our
clustering results. The majority of variables in the OGLE database are clearly
large-amplitude monoperiodic pulsators or binaries, but we do recover evidence
for the occurrence of some low-amplitude variables.

\subsection{The CoRoT intermediate archive}

In this section we show the results of the clustering analysis of the first four
runs of CoRoT exoplanet data (IRa01, LRa01, LRc01 and SRc01; see \citet{paper3}
for a description of the data). The goal of the exoplanet programme of the CoRoT
mission is to detect planets around stars through the transit method (see ``The
CoRoT Book'', \citet{corotbook}, for more explanation). In order to achieve
this, light curves of thousands of stars with a precision typically a factor 100
better than the Hipparcos and OGLE data have been gathered during long
uninterrupted sequences (5 months for a long run, a few weeks for a short run).
The probability of finding exoplanets is much higher for main-sequence stars
than for evolved stars.  As such, the CoRoT exoplanet database is heavily biased
towards main-sequence stars while giants and supergiants have been
avoided in the preselection of the targets as much as possible. It is therefore
evident that the population of variable stars in the CoRoT and OGLE databases
are almost opposite in nature, the Hipparcos variable star database bridging
these two extremes.

The analysis of the CoRoT dataset also differs from the previous two
ones in that no colour information was used in the analysis. The CoRoT
satellite is equipped with a dispersing prism that provides
information on colour changes with time for a given star, but the
masks are adapted to each of the stars separately and hence this
information cannot be translated into any kind of standard colour
index. Although there are ground-based colour indices available for a
fraction of the archive from the CoRoT EXODAT database
\citep{2007ASPC..376..339M}, we preferred not to use them for the
clustering because they have not been corrected for reddening. 
  This is also the case for the 2MASS \citep{2006AJ....131.1163S}
  colour indices. Although reddening effects should be less important
  in the infrared passbands, it has to be recalled that the CoRoT
  fields are located near the Galactic center (LRc01 and SRc01) and
  anticenter (IRa01, LRa01; see the 2MASS Explanatory Supplement to
  the Second Incremental Data Release for a comparison between typical
  colour-colour diagrams in Galactic plane and pole
  regions). Nevertheless, we have conducted a parallel experiment
  (described in section \ref{2masssect} below) in
  order to assess the robustness of our results when 2MASS colours are
  used as attributes for clustering.  Also, we would like the results
of the clustering analysis to be used directly in the update and
improvement of the supervised classification system which is based
solely on CoRoT light curves \citep{paper3}. The lack of colour
information will necessarily cause confusion among variability classes
whose light curves are very similar. This is the case, e.g., for
$\beta\,$Cep and $\delta\,$Sct stars, for SPBs and $\gamma\,$Dor
stars, and for various classes of stars with activity. We thus expect
these to be difficult to separate.

\begin{figure*}
   \centering
   \includegraphics[scale=0.5]{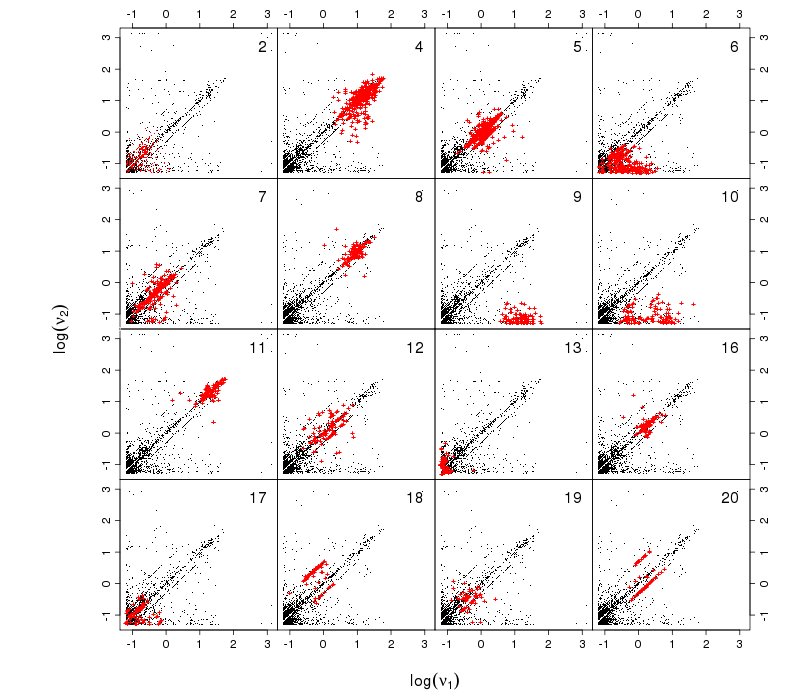}
   \caption{The clustering structure of the CoRoT first four runs
     archive at $\sigma=0.2$. The $x$-axis represents the logarithm of
     the first frequency and the $y$-axis, the logarithm of the second
     detected frequency. Black dots represent the complete
       database and red crosses identify cluster members.}
   \label{corot-nu1-nu2}
\end{figure*}

\begin{figure*}[htp]
   \centering
   \includegraphics[scale=.50]{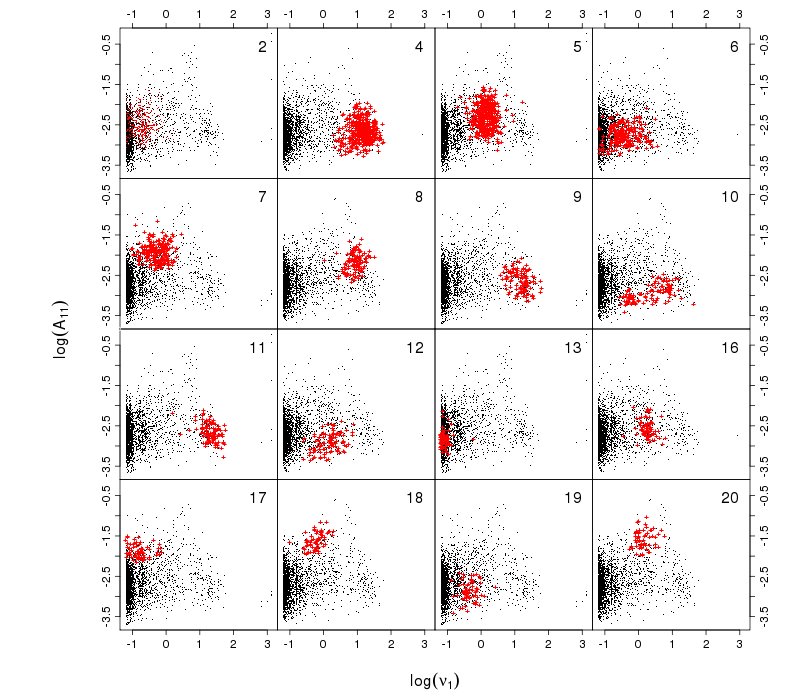}
   \caption{The clustering structure of the CoRot archive at
     $\sigma=0.2$. The $x$-axis represents the logarithm of the
     frequency and the $y$-axis, the logarithm of the amplitude
     $A_{11}$ in magnitudes of the first component in the Fourier
     decomposition for $\nu_1$. Black dots represent the complete
       database and red crosses identify cluster members.}
   \label{corot-nu1-a11}
\end{figure*}

On the other hand, the lack of colour information is partially
compensated for by the unprecedented quality of the spectral frequency
information that can be recovered from time series analysis of the
CoRoT exoplanet light curves.  For most of the light curves, numerous
frequencies at low amplitudes are detected and thus we may hope to
discriminate easily between monoperiodic and multiperiodic
variability. This is in contrast to the Hipparcos and OGLE light
curves, which quite often show only one significant frequency.  Hence,
for the CoRoT database, it is most instructive to look at the
clustering structure in the $(\nu_1, \nu_2)$ and $(\nu_1, A_{11})$
planes (Figs\,\ref{corot-nu1-nu2} and \ref{corot-nu1-a11}).  We
adapted the frequency threshold $\log (\nu_1) > -1.2$ as set in
\citet{paper3} in order to avoid long-term trends (of intrinsic or
instrumental nature) to dominate the frequency derivation. We clearly
see in Fig.\,\ref{corot-nu1-nu2} that many stars reside along the
bisectors in the subpanels, which points towards multiperiodic
variables. The most conspicuous OGLE clusters corresponding to the
monoperiodic RR~Lyrae stars, Cepheids, eclipsing binaries and small
amplitude red giants loci (more than 90\% of the OGLE variables) are
absent here. The implications for the supervised classification of
these targets is straightforward: the performance of a classifier
optimized for the OGLE sample of variable stars should not be
maintained when applied to such a different stellar population. This
is why the training set has been filled with CoRoT stars as much as
possible for the supervised classification in \citet{paper3}.

\begin{figure*}[H]
   \centering
   \includegraphics[scale=0.5]{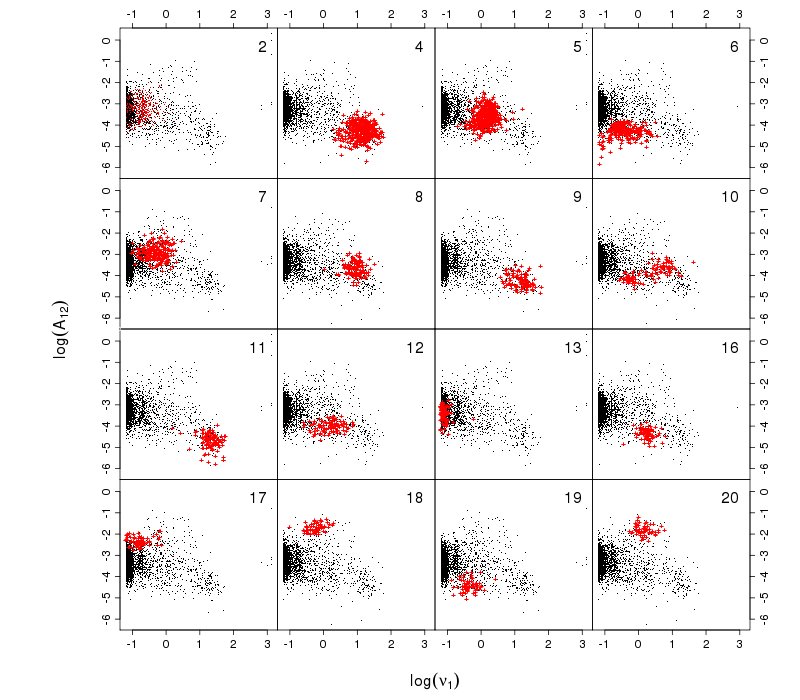}
   \caption{The clustering structure of the CoRoT first four runs
     archive at $\sigma=0.2$. The $x$-axis represents the logarithm of
     the first frequency and the $y$-axis, the logarithm of the
     amplitude of the second Fourier component of the first detected
     frequency. Black dots represent the complete database and
       red crosses identify cluster members.}
   \label{corot-nu1-a12}
\end{figure*}

\begin{figure*}[H]
   \centering
   \includegraphics[scale=0.5]{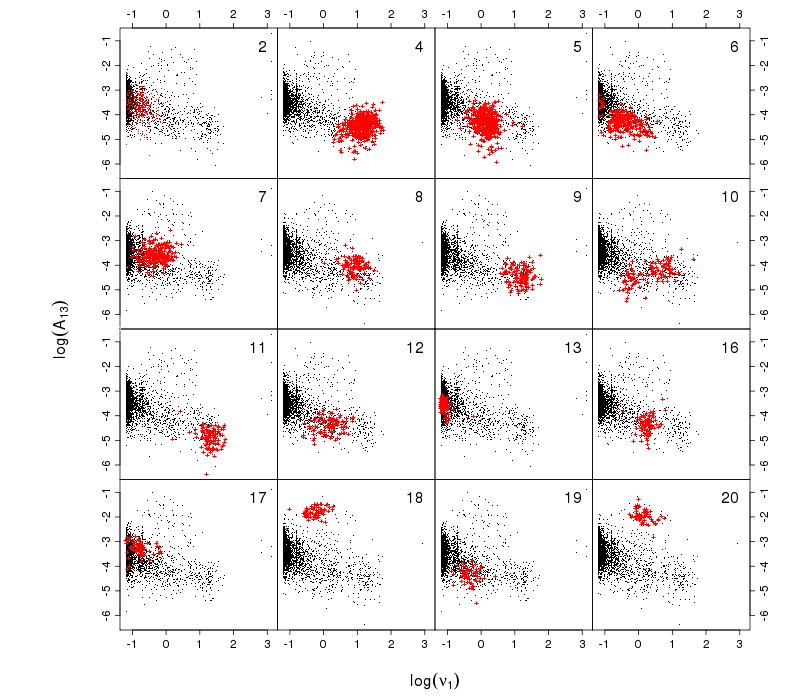}
   \caption{The clustering structure of the CoRoT first four runs
     archive at $\sigma=0.2$. The $x$-axis represents the logarithm of
     the first frequency and the $y$-axis, the logarithm of the
     amplitude of the third Fourier component of the first detected
     frequency. Black dots represent the complete database and
       red crosses identify cluster members.}
   \label{corot-nu1-a13}
\end{figure*}

\begin{figure*}[H]
   \centering
   \includegraphics[scale=0.5]{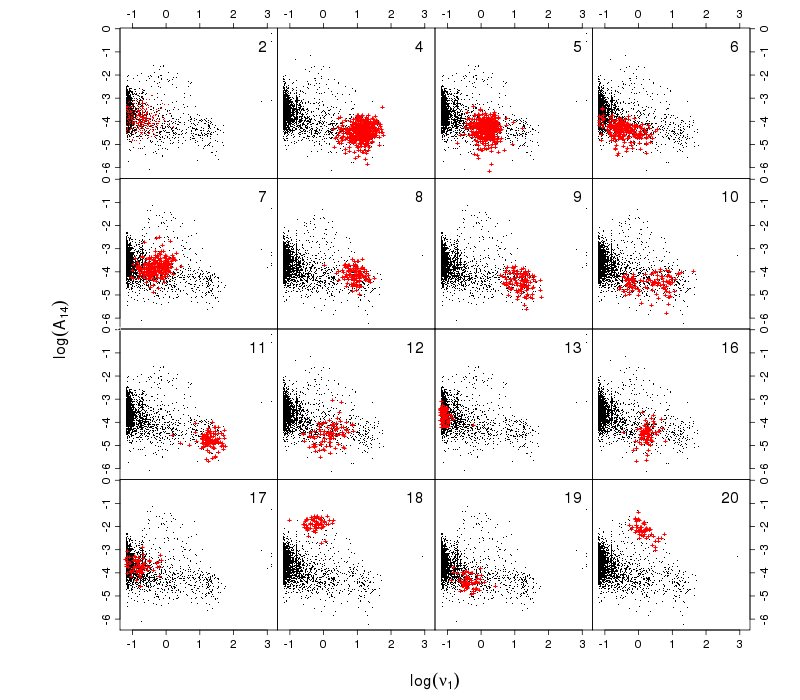}
   \caption{The clustering structure of the CoRoT first four runs
     archive at $\sigma=0.2$. The $x$-axis represents the logarithm of
     the first frequency and the $y$-axis, the logarithm of the
     amplitude of the fourth Fourier component of the first detected
     frequency. Black dots represent the complete database and
       red crosses identify cluster members.}
   \label{corot-nu1-a14}
\end{figure*}

\begin{figure*}[H]
   \centering
   \includegraphics[scale=0.5]{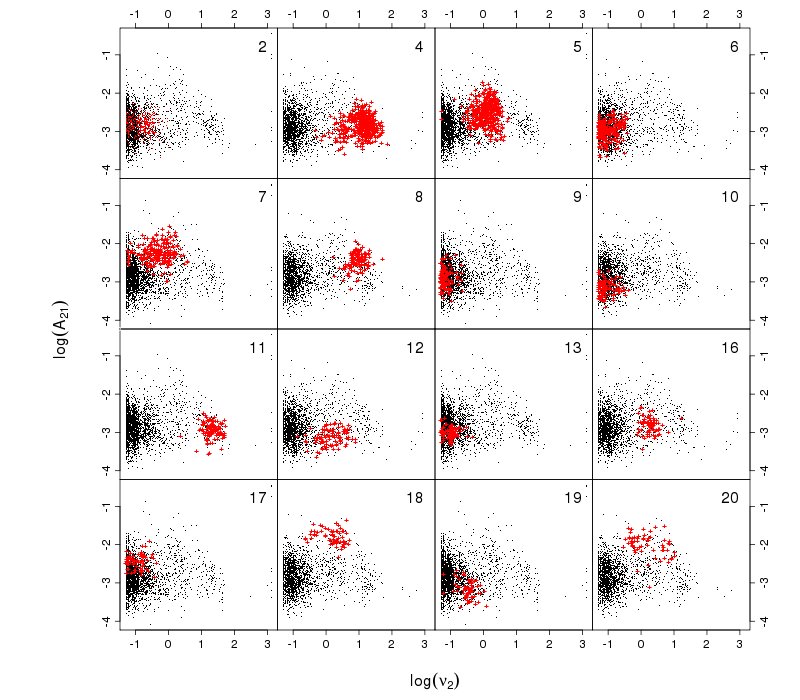}
   \caption{The clustering structure of the CoRoT first four runs
     archive at $\sigma=0.2$. The $x$-axis represents the logarithm of
     the second frequency and the $y$-axis, the logarithm of the
     amplitude of the first Fourier component of the second detected
     frequency. Black dots represent the complete database and
       red crosses identify cluster members.}
   \label{corot-nu2-a21}
\end{figure*}

\begin{figure*}[H]
   \centering
   \includegraphics[scale=0.5]{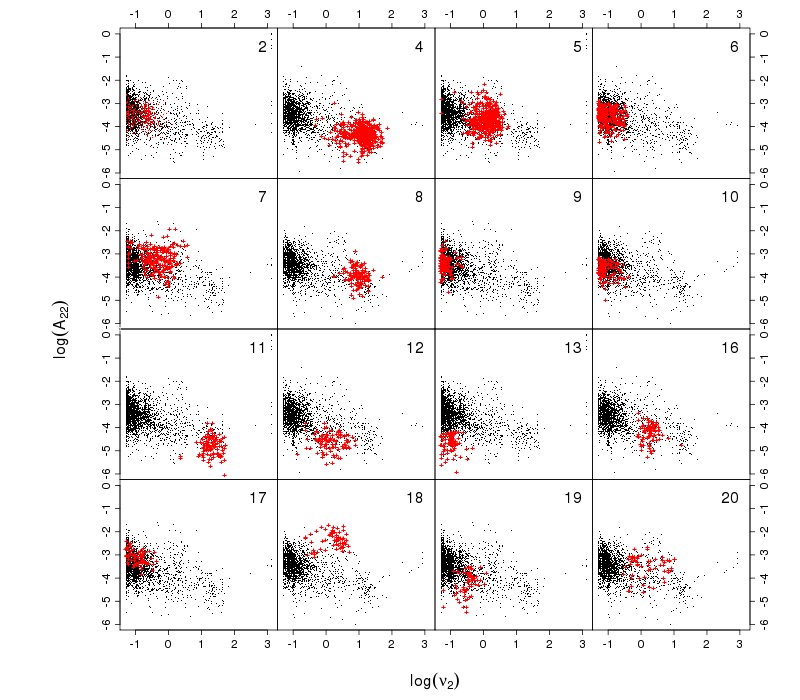}
   \caption{The clustering structure of the CoRoT first four runs
     archive at $\sigma=0.2$. The $x$-axis represents the logarithm of
     the second frequency and the $y$-axis, the logarithm of the
     amplitude of the second Fourier component of the second detected
     frequency. Black dots represent the complete database and
       red crosses identify cluster members.}
   \label{corot-nu2-a22}
\end{figure*}

\begin{figure*}[H]
   \centering
   \includegraphics[scale=0.5]{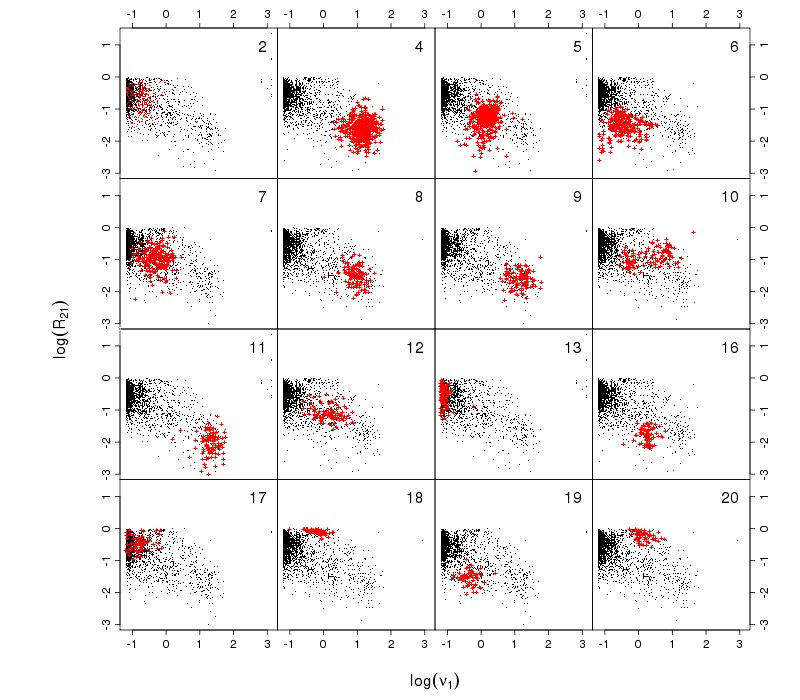}
   \caption{The clustering structure of the CoRoT first four runs
     archive at $\sigma=0.2$. The $x$-axis represents the logarithm of
     the first frequency and the $y$-axis, the logarithm of the
     $R_{21}$ ratio between the amplitudes of the first two components
     of the first frequency in the Fourier decomposition. Black
       dots represent the complete database and red crosses identify
       cluster members.}
   \label{corot-nu1-r21}
\end{figure*}

\begin{figure*}[H]
   \centering
   \includegraphics[scale=0.5]{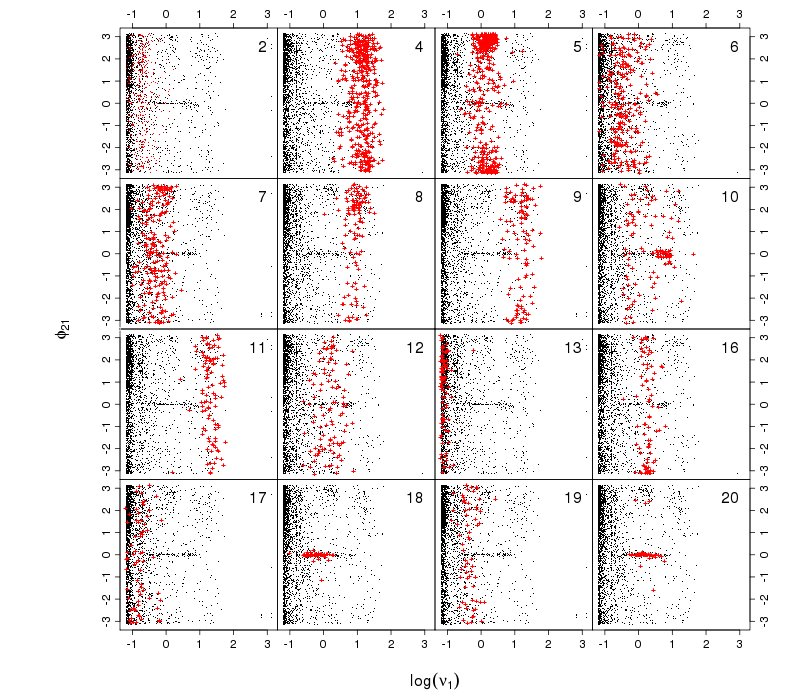}
   \caption{The clustering structure of the CoRoT first four runs
     archive at $\sigma=0.2$. The $x$-axis represents the logarithm of
     the first frequency and the $y$-axis, the phase difference
     $\phi_{12}$ between the first two components in the Fourier
     decomposition of the first frequency. The $\phi_{12}$ attribute
     was not used for clustering and it is only used here for
     illustration purposes. Black dots represent the complete
       database and red crosses identify cluster members.}
   \label{corot-nu1-fi21}
\end{figure*}

The interpretation of the clusters shown in Figs\,\ref{corot-nu1-nu2} and
\ref{corot-nu1-a11}, as well as in Figs\,\ref{corot-nu1-a12} ---
\ref{corot-nu1-fi21} is again based on their position in the various projections
of the parameter space, and also on the visual inspection of the first tens or
hundreds of time series and phase-folded light curves. As already emphasized, we
expect the interpretation of the clusters to be more difficult than for the
Hipparcos and OGLE cases, as the most prominent clusters for these databases are
absent here and as we expect groups of stars with similar periods to be mixed
due to lack of colour information.

The clusters shown in Figs\,\ref{corot-nu1-nu2} and \ref{corot-nu1-a11} are
sixteen of the largest clusters found, although many less populated yet still
significant clusters are found beyond these.  The reader will have noticed from
the subpanel labels that clusters 1, 3, 14 and 15 are absent. Visual inspection
of their light curves showed that these clusters contain mostly spurious
frequency detections due to jumps in the light curves as a consequence of hot
pixels spikes due to the passage of the satellite through the South Atlantic
Anomaly.  These effects are not intrinsic to the stars, which is why these
clusters are not discussed here.  For the time being we have no appropriate
solution to this problem, although automated jump correction and detrending
methods as in \citet{degroote2009} seem a promising route for future
applications before the clustering analysis.  We realise that these instrumental
artifacts sometimes occur together with underlying low-amplitude periodic signal
and that the latter unfortunately is masked by the jumps in the time series or
vice versa. We thus expect our retained clusters to be possibly contaminated by
light curves with small instrumental jumps, particularly for the clusters with
periodicities of the order of days.

Let us look at the low-amplitude multiperiodic oscillators first. As
already mentioned, we expect them to populate the bisector regions in
Fig.\,\ref{corot-nu1-nu2}, well above the point
$(\log\nu_1,\log\nu_2)=(-1,-1)$. This is the case for clusters 4, 5,
8, 11, 16, and less clearly also clusters 7, 12, and 19. Cluster 4, 8,
and 11 clearly cover the periodicities of the $\beta$ Cep/$\delta$ Sct
range, while clusters 5, 7, 12, and 16 correspond to frequencies in
the SPB/$\gamma$\,Dor range.  The dominant frequency of the cluster 9
stars is also $\delta\,$Sct-like. The class labels obtained with the
supervised classifiers presented in \cite{paper3} generally agree with
these cluster identifications based on their position in parameter
space.

\begin{figure*}[H]
   \centering
   \includegraphics[scale=0.5]{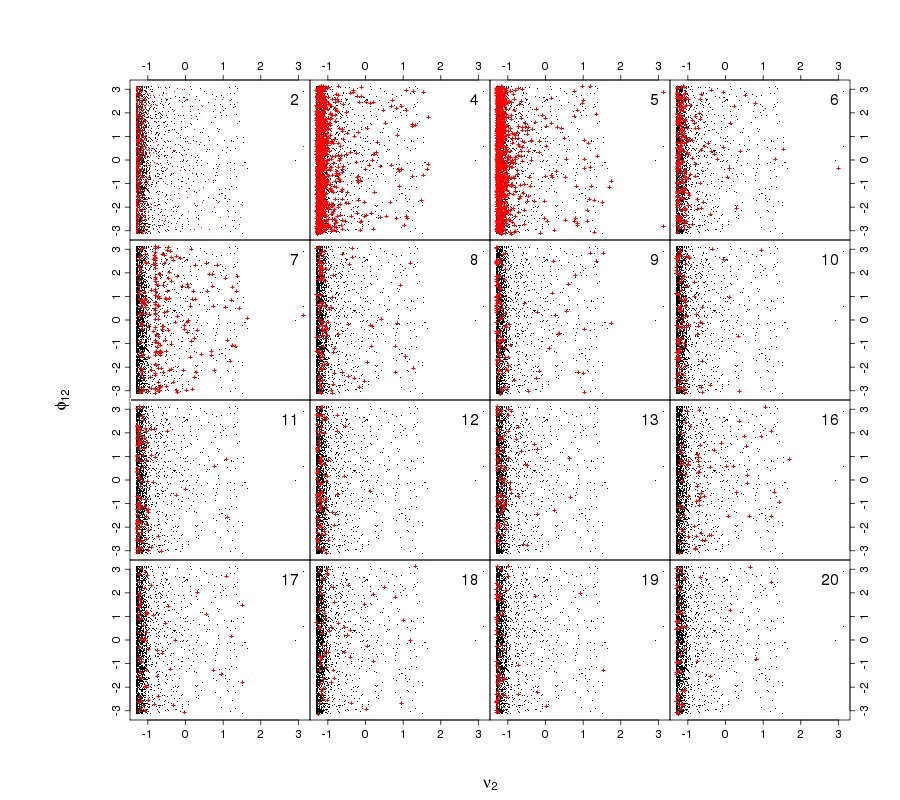}
   \caption{The clustering structure of the CoRoT first four runs
     archive at $\sigma=0.2$. The $x$-axis represents the logarithm of
     the second frequency and the $y$-axis, the phase difference
     $\phi_{12}$ between the first two components in the Fourier
     decomposition of this second frequency. The $\phi_{12}$ attribute
     was not used for clustering and it is only used here for
     illustration purposes. Black dots represent the complete
       database and red crosses identify cluster members.}
   \label{corot-nu2-fi21}
\end{figure*}

Based on the Initial Mass Function, $\delta$ Sct stars will
predominate in clusters 4, 8, 9, 11 over the $\beta$\,Cep population
(a hypothesis not contradicted by the unreddened EXODAT colours).
Clusters 4 and 9 gather the vast majority of $\delta$\,Sct stars (423
+ 112 stars) in the first four runs. The clusters 8 and 11 contain 117
and 103 stars respectively and they show distinctive features in
several projections of the parameter space. Cluster 8 is characterized
by lower frequencies, larger amplitudes and larger $R_{21}$ ratios
than clusters 4 and 9. We interprete this as the subgroup of
$\delta\,$Sct stars with nonlinear light curve distortion due to
high-amplitude (non)radial modes.  Cluster 11, on the contrary, seems
to contain the $\delta\,$Sct stars with extremely low amplitudes.
Cluster 9 is interpreted as the group of $\delta\,$Sct stars whose
second frequency is very low.  These are probably hybrid
$\delta\,$Sct/$\gamma\,$Dor stars, $\delta\,$Sct stars with rotational
modulation, or else $\delta\,$Sct stars in binaries with grazing
eclipses or ellipsoidal variability.  The latter two options are less
likely given that there is no specific phase relation for the second
frequency (Fig.\,\ref{corot-nu2-fi21}).  On the other hand, these
stars clearly have a higher value of $A_{22}$
(Fig.\,\ref{corot-nu2-a22}) and $R_{21}$ (Fig.\,\ref{corot-nu1-r21})
compared to those quantities for cluster 4 stars, pointing towards
deviations from sinusoidal light variability with frequency $\nu_2$.
This behaviour of $\nu_2$ can be interpreted in terms of rotational
modulation or of nonlinear g-mode oscillations triggered by resonant
mode coupling \citep{buchler}. It is very difficult to discriminate
between these two scenarios without spectroscopic information,
particularly since g modes may also be splitted into multiplets by
rotational effects on the oscillations.

\begin{figure*}[htp]
  \centering
  \includegraphics[scale=.50]{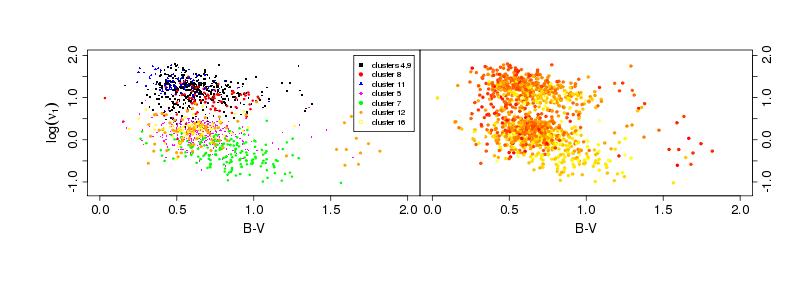}
  \caption{Frequency-colour diagrams of the low frequency
    multiperiodic pulsators clusters. The left plot shows clusters 4
    and 9 (black filled squares), 8 (red circles), 11 (blue
    triangles), 5 (magenta diamonds), 7 (green circles), 12 (filled
    orange) and 16 (open orange). The right panel shows the same plot
    with the logarithm of the amplitude of the first Fourier component
    colour coded from red (low amplitudes) to yellow (high
    amplitudes). The $x$ axis represents the EXODAT $B-V$ colour
    index, and the $y$ axis, the logarithm of the first detected
    frequency.}
  \label{colours}
\end{figure*}

\begin{figure*}[htp]
   \centering
   \includegraphics[scale=.50]{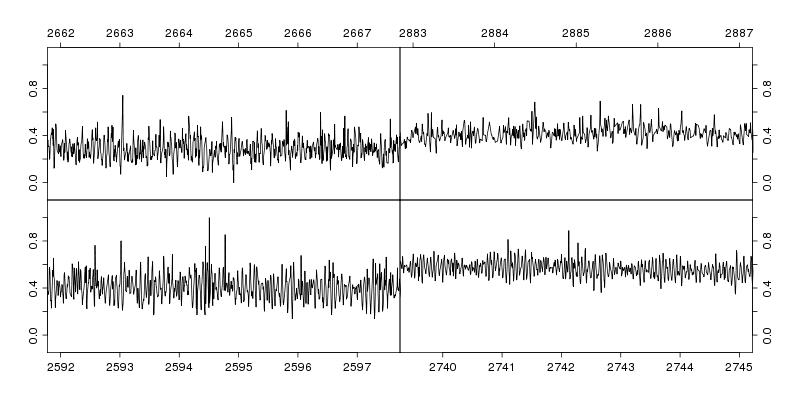}
   \caption{Close-up of the CoRoT photometric time series of the
     object closest to clusters 4 (top left), 8 (bottom left), 9
     (top right), and 11 (bottom right). The $x$ axis represents the
     modified heliocentric Julian Date, and the $y$ axis, the signal
     measured in counts.
}
\label{ex-4-8-9-11}
\end{figure*}

The clusters 8 and 11 show a clear bimodality in the unreddened
colour, $\nu_1$ diagram as shown in Fig.\,\ref{colours}, where the
stars of cluster 11 turn out to be somewhat bluer and those of cluster
8 redder than average.  This independent information clearly suggests
that there is a physical reason why these clusters were found as
separate classes, rather than this being due to statistical noise in
the point process. In fact, the low-amplitude $\delta\,$Sct stars in
cluster 11 seem to coincide with the low-amplitude pulsators filling
the gap between the classical instability strip and the SPB strip
found by \citet{degroote2009} which explains why they are somewhat
bluer than the majority of the $\delta\,$Sct stars.  On the other
hand, the stars in cluster 8 have clearly longer periods and are
redder, while their amplitudes are also somewhat higher than average.
This frequency--colour behaviour indicates that they are rather
evolved $\delta\,$Sct stars.  The time series of the objects closest
to the clusters modes are shown in Fig.\,\ref{ex-4-8-9-11} and fully
confirm our interpretation.

\begin{figure*}[htp]
   \centering
   \includegraphics[scale=.50]{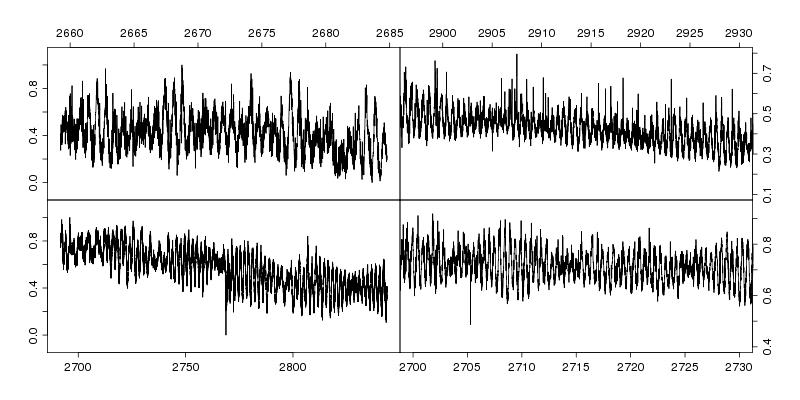}
   \caption{Close-up of the CoRoT photometric time series of the
     object closest to clusters 5 (top left), 7 (bottom left), 12 (top right), and 16 (bottom right). The $x$ axis
     represents the modified heliocentric Julian Date, and the $y$
     axis, the signal measured in counts.}
   \label{ex-5-16}
\end{figure*}

The differences amongst clusters 5, 7, 12, and 16 are mainly related
to the amplitudes of the first frequency. As shown in
Fig.\,\ref{corot-nu1-a11}, cluster 7 corresponds to the largest
amplitudes, cluster 5 stars also have relatively high amplitude while
cluster 12 and 16 shows significantly smaller amplitudes, and cluster
16 also somewhat higher frequencies.  Some of the light curves in the
vicinity of the mode of clusters 12 and 16 have low signal-to-noise
ratios. Moreover, cluster 16 is not only characterized by higher mean
frequencies, but also by lower mean values of $R_{21}$, which implies
that the oscillations in that group of stars is highly linear, while
modest nonlinear effects seem to occur for the stars in cluster 5, 7,
and 12.  All these properties suggest that clusters 5, 7, 12, and 16
contain SPB and $\gamma\,$Dor stars, which are both classes of
multiperiodic gravity-mode nonradial oscillators.  The time series of
the four objects closest to the cluster 5 (top left), 7 (bottom left),
12 (top right) and cluster 16 (bottom right) modes are shown in
Fig.\,\ref{ex-5-16} and support this interpretation.

The position of the stars in these four clusters in the unreddened
colour diagram shown in Fig.\,\ref{colours} seems to suggest that the
clusters 5, 12, and 16 contain a mixture of SPB and $\gamma\,$Dor
stars, with the majority being $\gamma\,$Dor stars, again based on the
Initial Mass Function. The stars have essentially the same colour
properties, which confirms that the bluer SPBs must be relatively low
in number than the redder $\gamma\,$Dor stars.  The stars in cluster 7
are clearly redder and have longer periods connected with their higher
amplitudes.  Again, we interprete this as an evolutionary effect and
thus this cluster contains the evolved $\gamma\,$Dor stars. A striking
feature in Fig.\,\ref{colours} are some ten stars of cluster 12 and
one of cluster 7 with far redder colours than all others in the four
clusters with g-mode pulsators.  While we have to be careful for
overinterpretation of Fig.\,\ref{colours} due to lack of an
appropriate correction for reddening effects, it seems to suggest that
these stars are much more evolved than the other g-mode pulsators.  It
could be that this small group of stars represent the PVSG class,
which are evolved B stars with SPB like oscillations with longer
periods \citep{lefever}.

We cannot exclude that both cluster 8 and 7 are contaminated by a few
pre-main-sequence $\delta\,$Sct and $\gamma\,$Dor stars, respectively,
which would still be surrounded by remnant material of their birth
cloud and which would be an alternative explanation for their redder
colour.

Finally, we point out that the frequency -- colour behaviour displayed in
Fig.\,\ref{colours} is in full agreement with the theoretical predictions of the
instability strips of the $\delta\,$Sct and $\gamma\,$Dor stars, along with a
small number of pulsating B stars, as presented in, e.g., \citet{degroote2009}.

\begin{figure}[htp]
   \centering
   \includegraphics[scale=.30]{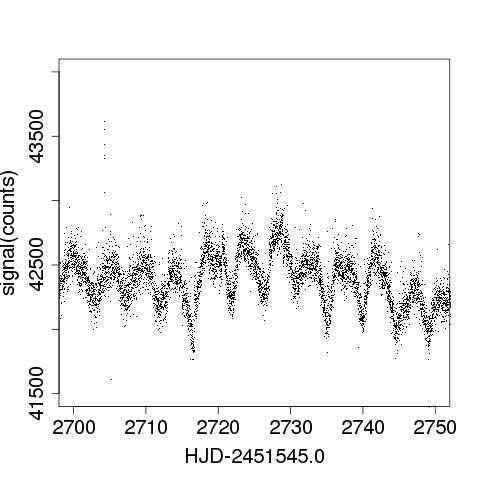}
   \caption{Close-up of the CoRoT photometric time series of the
     object closest to the cluster 2 mode. The $x$ axis represents the modified
     heliocentric Julian Date, and the $y$ axis, the signal measured
     in counts.}
   \label{corot-activity}
\end{figure}

Several clusters contain stars whose variability patterns we interprete in terms
of phenomena related to stellar activity.  The largest cluster (cluster 2) is
composed (again, according to the time series of the first tens of objects
closest to the cluster mode) of stars showing time series characterized by the
occurrence of harmonics of the dominant frequencies, which points to strongly
non-sinusoidal light curves. Often, this can be explained in terms of one or
several starspots, that can migrate in phase with respect to one another. A
typical light curve corresponding to the objects closest to the cluster mode is
shown in Fig.\,\ref{corot-activity}.

\begin{figure*}[htp]
   \centering
   \includegraphics[scale=.30]{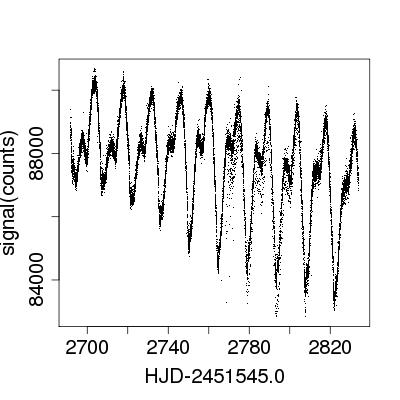}
   \caption{Close-up of the CoRoT photometric time series of the
     object closest to the cluster 17 mode. The $x$ axis represents
     the modified heliocentric Julian Date, and the $y$ axis, the
     signal measured in counts.}
   \label{camel}
\end{figure*}

Cluster 17 contains the most extreme examples of strict rotational modulation
active in the stellar photospheres, in the sense of being characterized by some
of the largest amplitudes in the CoRoT sample as well as very high
$R_{21}$-values (see Figs\,\ref{corot-nu1-r21} and \ref{camel}).

\begin{figure*}[htp]
   \centering
   \includegraphics[scale=.30]{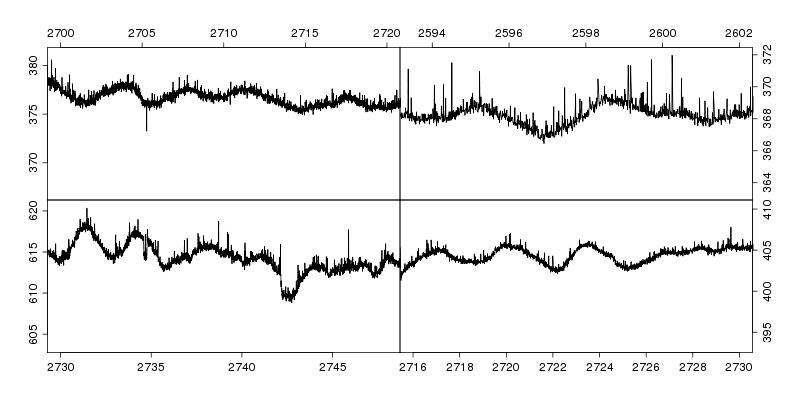}
   \caption{Close-up of the CoRoT photometric time series of the
     objects closest to the cluster 6 mode. The $x$ axis represents
     the modified heliocentric Julian Date, and the $y$ axis, the
     signal measured in units of $10^3$ counts.}
   \label{ex-6}
\end{figure*}

Clusters 6 and 10 show similar behaviour but with an interesting peculiarity:
the variability of these stars is systematically described by two very different
dominant frequencies (see Fig.\,\ref{corot-nu1-nu2}). Cluster 6 shows clear
signs of bimodality, something that is confirmed by clustering analysis with
smaller kernel bandwidths: it is composed of two populations, one with similar
values of the first two detected frequencies, interpreted as complex activity,
and one characterized by values of $\log(\nu_1)$ between -1 and 1, and typical
values of $\log(\nu_2)$ less than -1.0. The latter stars are interpreted as
active stars whose light curves also show long-term trends.  Fig.\,\ref{ex-6}
shows close-ups of time series in the vicinity of the modes of cluster 6. We can
see very conspicuous low frequency modulations superposed to an activity
signature.

\begin{figure*}[htp]
   \centering
   \includegraphics[scale=.30]{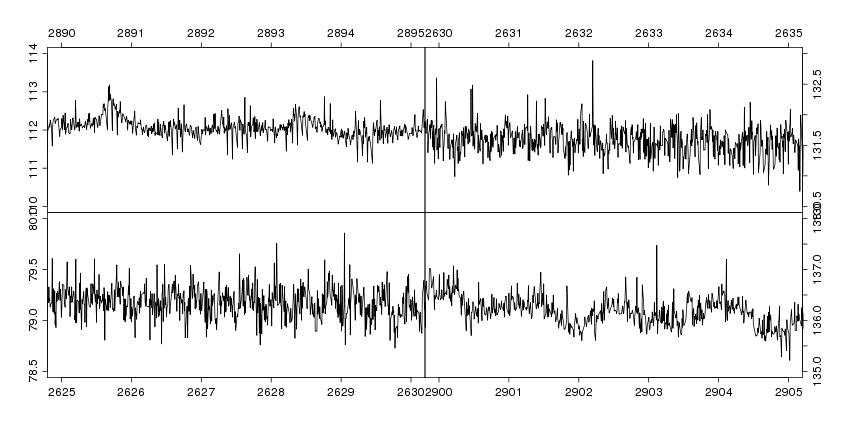}
   \caption{Close-up of the CoRoT photometric time series of the
     objects closest to the cluster 10 mode. The $x$ axis represents
     the modified heliocentric Julian Date, and the $y$ axis, the
     signal measured in units of $10^3$ counts.}
   \label{ex-10}
\end{figure*}

Cluster 10 is also bimodal at smaller kernel bandwidths, but both subclusters
are characterized by $\log(\nu_2)$ values around -1.0, i.e., a long-term trends,
while being bimodal in their $\log(\nu_1)$, with values around -0.15 and 0.8,
respectively. It may very well be that this groups consists of pulsating Be
stars already discussed above in the framework of the Hipparcos database. These
stars indeed undergo trends due to some level of activity and/or outbursts,
while showing also oscillations with frequencies of pressure modes or gravity
modes.  Fig.\,\ref{ex-10} show close-ups of time series in the vicinity of the
modes of cluster 10.  These light curves are free from significant jumps. It has
to be beared in mind that, in these clusters, it is the high frequency component
that is detected first in the power spectrum, and it is thus associated to a
higher peak.  Future spectroscopic data are needed to refine this preliminary
interpretation of cluster 10.

\begin{figure*}[htp]
   \centering
   \includegraphics[scale=.30]{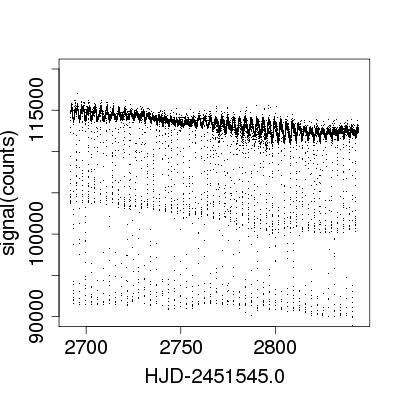}
   \caption{Close-up of the CoRoT photometric time series of the
     object closest to the cluster 18 mode (eclipsing binaries). The
     $x$ axis represents the modified heliocentric Julian Date, and
     the $y$ axis, the signal measured in counts.}
   \label{ex-18}
\end{figure*}

The majority of eclipsing binaries are collected in clusters 18, 20
(and 21, and 22, plots not shown).  This can be readily deduced from
the phase behaviour of the dominant frequency as illustrated in
Fig.\,\ref{corot-nu1-fi21}.  It turns out that the object closest to
the mode of cluster 18 is a binary system (or a blend within the mask
of CoRoT) with a multiperiodic component (see Fig.\,\ref{ex-18}). In
general, we found a few eclipsing binaries with pulsating components
in the clusters with pulsators discussed above. The majority of stars
in the four binary clusters 18, 20, 21, 22 only show eclipses,
although some also have a signature of rotational modulation and/or
activity outside the eclipses.

Finally, cluster 19 is a mixture of binaries, stars with activity, and some
seemingly multiperiodic stars, while cluster 13 contains a mixture of stars with
artificial jumps.

\subsubsection{\label{2masssect} Clustering with 2MASS colours}

\begin{figure*}[htp]
   \centering
   \includegraphics[scale=.50]{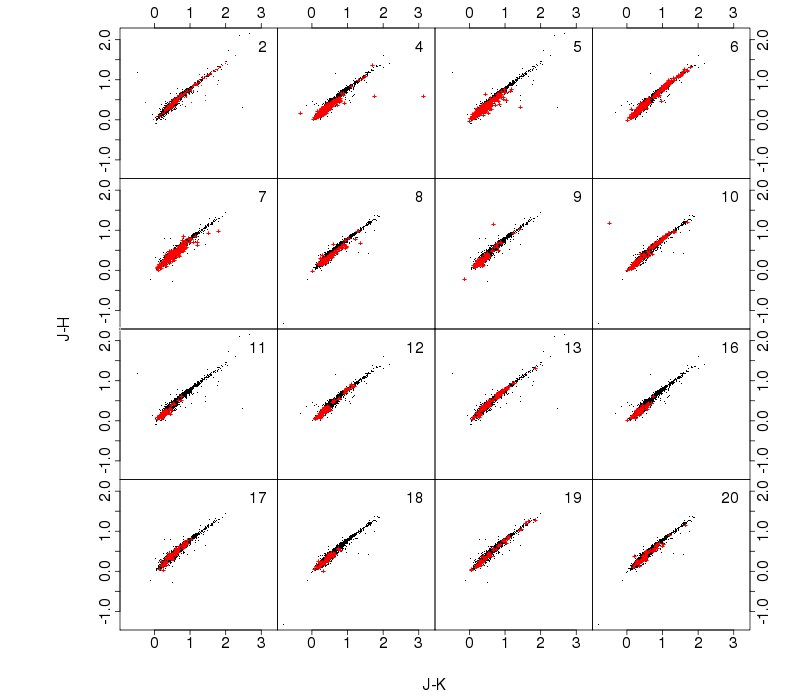}
   \caption{2MASS colour-colour diagrams of the CoRot clusters
     obtained with $\sigma=0.2$ and no colour index used as
     atribute. The $x$ and $y$-axis represent the $J-K$ and $J-H$
     colour indices respectively.  Black dots represent the
       entire sample and red symbols correspond to cluster members.}
   \label{2mass1}
\end{figure*}

We have conducted two complementary experiments in order to evaluate
the information that the 2MASS colour indices could potentially
provide to the clustering process. First, we analysed the clustering
structure described above in the light of the infrared properties of
the CoRoT targets (i.e., we retain the clusters found above and
analyse the characteristic colours of the stars in these
clusters). Figure \ref{2mass1} shows the colour-colour diagram of the
CoRoT sample, with the cluster membership described in the
previous section superimposed as red symbols.

\begin{figure*}[htp]
   \centering
   \includegraphics[scale=.50]{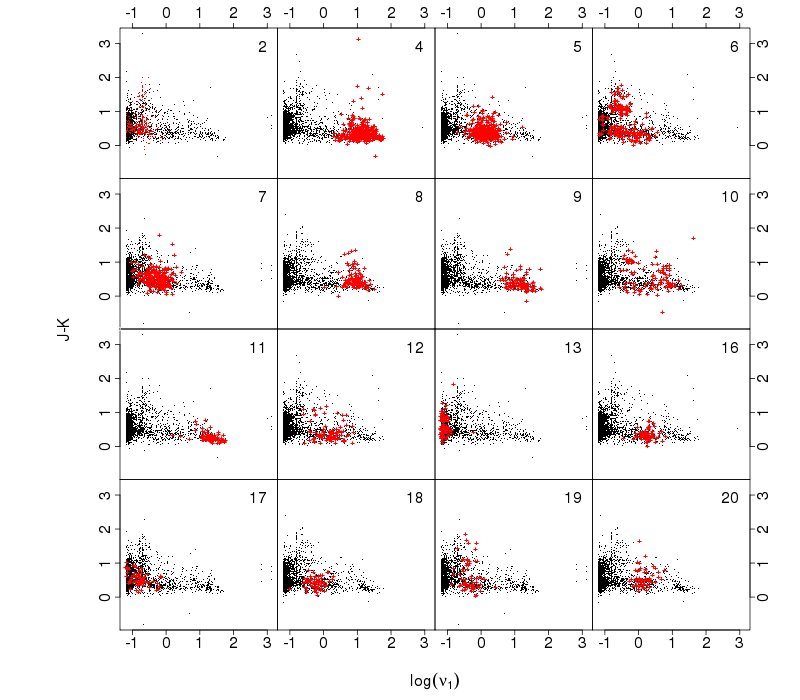}
   \caption{The clustering structure of the CoRoT first four runs
     archive at $\sigma=0.2$. The $x$-axis represents the logarithm of
     the first frequency and the $y$-axis, the $J-K$ colour
     index. Black dots represent the entire sample and red
       symbols correspond to cluster members.}
   \label{2mass2}
\end{figure*}

It is evident from the plots that the reddening has smeared out the
underlying distribution (see the Galactic pole colour-colour diagram
in the Explanatory Supplement to the 2MASS Second
Release). Nevertheless, Fig.\,\ref{2mass2} shows some interesting
features that arise when the 2MASS colour indices are interpreted in
the light of the frequency content of the light curves. First of all,
contamination of cluster 2 by spurious frequency detections due to
discontinuities in the time series is apparent as vertical lines in
the corresponding plot. Second, the bimodality in frequency space of
clusters 6 and 10 (described above) seems to correlate well with the
$J-H$ colour index, in the sense that the subgroups with smaller
frequencies correspond partly to a distinct group of redder,
most probably giant stars.

In the second experiment, we carried out a new clustering of the CoRoT
database, but this time we extended it with the 2MASS colour
indices. We then studied the new groups of stars in relation with the
clusters obtained without 2MASS data. The results can be summarized as
follows:
\begin{itemize}
\item The clusters interpreted as $\delta\,$ Scuti and $\beta\,$
  Cephei stars (4, 8, 9, and 11) remain as easily recognizable
  clusters in the new clustering structure. Cluster 9 is split into
  two clusters, and the other three are preserved, with a clear
  tendency for cluster 4 to absorb a significant fraction of clusters
  8 and 11. The original differences between these clusters (described
  thoroughly in the previous section) seem to be smeared out by the
  use of the 2MASS colours. Little or no contamination is observed
  from other clusters;
\item stars that belonged to the clusters interpreted as $\gamma\,$
  Doradus and Slowly Pulsating B stars (5, 7, 12, and 16) remain
  grouped in a number of clearly identifiable clusters, and are not
  contaminated by other clusters. There is again a tendency for the
  clusters to merge into one larger cluster, except for cluster 7 that
  remains largely separated;
\item clusters numbered 1, 2 and 3 in the clustering experiment
  without 2MASS colours, remain the three largest groups when the
  latter are taken into account, but the contigency table (the table
  that summarizes the number of stars previously in cluster $i$ and
  now in cluster $j$) shows a major redistribution of stars in these
  categories. We interpret this as a hint that objects in clusters 1
  and 3 should not be automatically disregarded as
  ``artifacts''. Further studies are needed to better separate these
  from real signals of astrophysical origin;
\item the clearly bimodal cluster 6 splits into two larger clusters
  (interpreted as dwarves and giants in the paragraph above) and two
  smaller ones; a similar behaviour is observed in cluster 10, with
  some stars merging into one of the smaller clusters originated in
  cluster 6.
\item finally, the eclipsing binaries remain clearly separated as
  groups, with minor redistributions of stars in clusters. 
\end{itemize}

\section{Conclusions}
\label{conclusions}

We presented the results of extensive clustering experiments with three
variability databases of very different characteristics, using a density based
approach. We concentrated on the large scale clustering properties of the
samples, trying to understand what domain of the variability zoo can be
recovered with this kind of multivariate analysis. While the Hipparcos archive
is characterized by a small number of objects with reliable class assignments
which simplify the interpretation of the clusters, the CoRoT database is only
starting to be understood due to the very high level of precision.

The results presented here already constitute a clear cut separation of
variability types, as we understand them based on theoretical grounds. The
clusters discussed above have been interpreted on the basis of their average
properties, and on the visual inspection of those objects close to the clusters
modes. Since these modes are located at the regions of maximum probability
density, these are the prototypes of the clusters, but significant contamination
from a variety of other objects cannot be discarded, and should, in fact, be
assumed. This is mainly the case for those types where the attributes used to
describe the objects are insufficient for a unique description of the time
series. Nevertheless, the existence of well defined clusters containing several
types of light curves with the imprint of nonradial oscillations, rotational
modulation, activity, binarity, etc., in the CoRoT database, encourages the use
and improvement of the current attribute set to interprete the detailed physics
of the clusters.

We used the Hipparcos and OGLE LMC archives in order to define a reference frame
that helps both the interpretation of the CoRoT database clustering structure
and the discovery of new classes in it.  In the Hipparcos archives, we clearly
managed to separate the classical pulsators (Cepheids and RR~Lyrae stars) from
the eclipsing binaries and from the multiperiodic low-amplitude pulsators along
the main sequence. This was successful thanks to the availability of a
well-calibrated colour index.  In the LMC case, the use of reddening free
attributes like the Wessenheit index allows for the recovery of the various
sequences of red giants in the long period range of the archive. Furthermore,
the ellipsoidal variables are easily recognized in the OGLE clusters thanks to
the $\log(\nu)$--$W_I$ correlation that they follow.

For the CoRoT exoplanet database, we have shown that we are capable of
discovering various different properties among the low-amplitude multiperiodic
stars.  Although the confirmation of the various proposed subclasses among
known nonradial pulsators and stars with activity must await the collection of
complementary information, likely in the form of spectra, we are capable of
discovering refinements of the properties of globally understood variability.
Such refinements are only possible thanks to large scale clusters unravelled in
the data.

The unprecedented quality of the time series provided by CoRoT has opened a
completely new realm of variability, where low amplitude signals constitute new
clusters by themselves, or appear as second or third frequencies combined with
large amplitudes ones, sometimes giving rise to a separate cluster.

One of the most interesting findings, is the clear separability of the
subpopulations among the previously known classes nonradial pulsators along the
main sequence.  Prior to the CoRoT mission, we had only been able to hint at the
properties of larger versus lower amplitude members of the real distribution of
these objects in paramater space, as well as on the suspicion of hybrid
pulsators showing pressure and gravity modes simultaneously, for the
$\beta\,$Cep/Be versus SPB classes on the one hand, and for the $\delta\,$Sct
versus $\gamma\,$Dor stars on the other hand.  Also, the clustering pointed
towards pulsating stars of various kinds in close binaries.  Completely new
sample of these pulsators will soon be publicly available to the scientific
community for further studies, to be complemented with other sources of
information.

All this knowledge gained with the unsupervised classification of the CoRoT
database needs to be incorporated in the framework of the supervised
classifiers, especially in the case of the light curve characteristics due to
rotational modulation and stellar activity. These variables turn out to
represent a very significant fraction of the database while they are not
sufficiently well represented in the training set used for the supervised
classification \citep{paper3}.

A systematic analysis of the smallest clusters as well as the
decomposition into finer scales is left for future work.  In the case
of the CoRoT archive, this further analysis will be complemented with
data from new runs, enhancing seriously the density clumps at the base
of the method. This will largely facilitate the discovery of potential
new (sub)classes of variability. We also stress that improvements in
the handling and removal of the instrumental artifacts in the CoRoT
photometric time series, which occurred for a larger number of objects
in the presented samples, will reveal other potential new types of
variability not found here and will hopefully seriously improve the
already very good performance of our algorithm.

Finally, even though the main aim of this work was the preparation for the
analysis of the CoRoT database, we believe that the results found in
the LMC dataset justify a comparative analysis of the OGLE Galactic
Bulge data using the same approach.

\begin{acknowledgements}
The research leading to these results has received funding from the Spanish
Ministerio de Educaci\'on y Ciencia through grant AYA2005-04286, from the
European Research Council under the European Community's Seventh Framework
Programme (FP7/2007--2013)/ERC grant agreement n$^\circ$227224 (PROSPERITY),
from the Belgian PRODEX programme under grant PEA C90199 (CoRoT Mission Data
Exploitation II), and from the research Council of Leuven University under grant
GOA/2008/04. The authors made use of the Spanish Virtual Observatory supported
from the Spanish MEC through grants AyA2005-04286, AyA2005-24102-E.
\end{acknowledgements}

\bibliographystyle{aa}
\bibliography{sarro}

\Online

\clearpage

\clearpage

\end{document}